\documentclass[12pt,preprint]{aastex}
\usepackage{psfig} 

\def\kms{~km~s$^{-1}$}
\def\hal{H$\alpha$}
\def\be{\begin{equation}}
\def\ee{\end{equation}}

\def\m{~$\mu$m}
\def\HI{\ion{H}{1}}
\def\HII{\ion{H}{2}}

\def\NII{[\ion{N}{2}]}

\def\CII{[\ion{C}{2}]}

\def\OII{[\ion{O}{2}]}
\def\OIII{[\ion{O}{3}]}

\def\Pab{Pa$\beta$}
\def\Brg{Br$\gamma$}
\def\Htwo{H$_2$}
\def\FeII{[\ion{Fe}{2}]}
\def\FeIIa{[\ion{Fe}{2}]1.257\m}
\def\FeIIb{[\ion{Fe}{2}]1.644\m}

\def\IRAScolor{${f_\nu (60 \mu {\rm m})} \over {f_\nu (100 \mu {\rm m})}$}

\def\colore{$f_\nu(6.75 \mu {\rm m})/f_\nu(15 \mu {\rm m})$}
\def\COa{$^{12}$CO($1\rightarrow0$)}

\def\COc{$^{13}$CO($1\rightarrow0$)}

\begin {document}
\slugcomment{\scriptsize \today \hskip 0.2in Version 2.3}

\title{Near-Infrared Integral Field Spectroscopy of Star-Forming Galaxies}
 
\author{Daniel A. Dale,\altaffilmark{1} H\'el\`ene Roussel,\altaffilmark{2} Alessandra Contursi,\altaffilmark{3} George Helou,\altaffilmark{2,4} Harriet L. Dinerstein,\altaffilmark{5} Deidre A. Hunter,\altaffilmark{6} David J. Hollenbach,\altaffilmark{7} Eiichi Egami,\altaffilmark{8} Keith Matthews,\altaffilmark{2} Thomas~W.~Murphy,~Jr.,\altaffilmark{9} Christine E. Lafon,\altaffilmark{10} and Robert H. Rubin\altaffilmark{7}}
\altaffiltext{1}{\scriptsize Department of Physics and Astronomy, University of Wyoming, Laramie, WY 82071; ddale@uwyo.edu}
\altaffiltext{2}{\scriptsize California Institute of Technology, Pasadena, CA 91125}
\altaffiltext{3}{\scriptsize Max-Plank-Institut f\"{u}r extraterrestrische Physik, Giessenbachstrasse 85748 Garching, Germany}
\altaffiltext{4}{\scriptsize Infrared Processing and Analysis Center, Pasadena, CA 91125}
\altaffiltext{5}{\scriptsize Department of Astronomy, University of Texas, RLM 15.308, Austin, TX 78712}
\altaffiltext{6}{\scriptsize Lowell Observatory, 1400 Mars Hill Road, Flagstaff, AZ 86001}
\altaffiltext{7}{\scriptsize NASA/Ames Research Center, MS 245-6, Moffett Field, CA 94035}
\altaffiltext{8}{\scriptsize Steward Observatory, University of Arizona, 933 North Cherry Avenue, Tucson, AZ 85721}
\altaffiltext{9}{\scriptsize Department of Physics, University of Washington, Box 351560, Seattle, WA 98195}
\altaffiltext{10}{\scriptsize Harvard-Smithsonian Center for Astrophysics, 60 Garden Street, Cambridge, MA 02138}

\begin {abstract}
The Palomar Integral Field Spectrograph was used to probe a variety of environments in nine nearby galaxies that span a range of morphological types, luminosities, metallicities, and infrared-to-blue ratios.  For the first time, near-infrared spectroscopy was obtained for nuclear or bright \HII\ regions in star-forming galaxies over two spatial dimensions (5\farcs7~$\times$~10\farcs0) in the \FeII~(1.257\m), \FeII~(1.644\m), \Pab~(1.282\m), \Htwo~(2.122\m), and \Brg~(2.166\m) transition lines.  These data yield constraints on various characteristics of the star-forming episodes in these regions, including their strength, maturity, spatial variability, and extinction.  The \HII\ regions stand out from the nuclei.  Unlike observations of nuclear regions, \HII\ region near-infrared observations do not show a spatial coincidence of the line and continuum emission; the continuum and line maps of \HII\ regions usually show distinct and sometimes spatially-separated morphologies.  Gauging from \Pab\ and \Brg\ equivalent widths and luminosities, the \HII\ regions have younger episodes of star formation than the nuclei and more intense radiation fields.  Near-infrared line ratio diagnostics suggest that \HII\ regions have ``purer'' starbursting properties.  The correlation between ionizing photon density and mid-infrared color is consistent with the star formation activity level being higher for \HII\ regions than for nuclei.  And though the interpretation is complicated, on a purely empirical basis the \HII\ regions show lower Fe$^{1+}$ abundances than nuclei by an order of magnitude.
\end {abstract}
 

\section {INTRODUCTION}
Stellar processing is the main evolutionary process affecting the cosmos.  Nucleosynthesis and mass return to the interstellar medium modify the chemical composition of the Universe and the fraction of mass bound in stars.  Star formation dominates the present-day radiation field, and most of it happens in normal galaxies (e.g. Kim \& Sanders 1998; Driver 1999).  Relatively little, however, is known about star formation on the scale of a galaxy, including its drivers and inhibitors, and the role of interactions (e.g. Volker 2000; Dopita et al. 2002; Hameed \& Young 2003).  The $ISO$ Key Project on the Interstellar Medium of Normal Galaxies (Helou et al. 1996; Dale et al. 2000) aimed to better understand the large-scale physics of the interstellar medium through an array of mid- and far-infrared data on a diverse sample of 69 nearby star-forming galaxies (e.g. Dale et al. 1999, Dale et al. 2000, Helou et al. 2000, Hunter et al. 2001, Malhotra et al. 2001, Contursi et al. 2002, Lu et al. 2003).  In this contribution we present and discuss near-infrared integral field spectroscopy for several of these objects.

For optical depths $A_V \gtrsim 3$, the brightest emission lines that probe the star-forming interstellar medium are found in the near-infrared.  Vibrationally excited \Htwo\ emission at 2.122\m\ arises by collisional excitation from shocks or radiative excitation in intense ultraviolet radiation fields (e.g. Goldader et al. 1997), or possibly from reprocessed X-ray illumination (Maloney, Hollenbach, \& Tielens 1996).  On the other hand, the \FeII\ lines at 1.257 or 1.644\m\ probe supernova-shocked gas (the final stage of processing) and hard power-law/X-ray environments, both of which can release iron atoms and ions through interstellar dust grain sputtering and evaporation; recent $HST$ work on M~82 and NGC~253 by Alonso-Herrero et al. (2003) indicates that as much as 70\% of the \FeIIb\ flux ultimately derives from supernovae, and only 6--8\% from \HII\ regions.  In contrast to \FeII, hydrogen lines like \Pab~(1.282\m) and \Brg~(2.166\m) directly trace the gas ionized by young massive stars.  Thus, an \FeII-to-hydrogen line ratio can be interpreted as an indicator of the maturity of the local star formation episode, as reflected in the ratio of supernova-shocked gas to molecular gas feeding star formation.  Moreover, the ratios \FeIIa/\Pab\ and \Htwo(2.122\m)/\Brg\ discriminate between shock excitation from supernova remnants or hard X-ray heating from power-law sources (Larkin et al. 1998), and are essentially unaffected by reddening.  

In addition to studying these diagnostics, we use the near-infrared line and continuum images to explore the maturity and spatial progression in the star formation.  Coupled with \hal\ imaging, the near-infrared hydrogen line fluxes are used to estimate the optical depth and reconstruct the intrinsic emission, total ionizing flux, and other properties of the local interstellar medium.  Data at other wavelengths help to further probe the interstellar medium.  For example, new optical spectroscopy and archival mid-infrared imaging allow us to investigate trends with metallicity and infrared colors.  The various results will point to important physical differences between nuclear and extranuclear/\HII\ regions in star-forming galaxies.

\section{THE SAMPLE}
The targets (Tables~\ref{tab:sample} and \ref{tab:obs} and Figures~\ref{fig:2mass1}--\ref{fig:2mass5}) derive from the $ISO$ Key Project on the Interstellar Medium of Normal Galaxies.  The Key Project collected infrared data for 69 star-forming galaxies:  carbon, oxygen, and nitrogen fine-structure line fluxes between 50 and 160\m\ (Malhotra et al. 2001), mid-infrared spectra between 3 and 12\m\ (Lu et al. 2003), and mid-infrared maps at 7 and 15\m\ (Dale et al. 2000).  In addition, the following ancillary ground-based data have been obtained for a large portion of the sample: broadband $B$ and narrowband \hal\ imaging, long-slit optical spectroscopy, and literature CO fluxes.   

The Key Project sample had the original following criteria: $f_\nu(60$ \m) $\gtrsim 3$ Jy; a published redshift; and no AGN or Seyfert classification on NED (at that time).  The sample explores the full range of morphology, luminosity ($L_{\rm FIR}$ from less than $10^{8} L_\odot$ to as large as $10^{12} L_\odot$), infrared-to-blue ratio (0.05 to 50) and infrared colors (see \S~2 of Dale et al. 2001 for a complete description of the sample selection).  The subset of the Key Project sample selected for this particular ground-based follow-up project was chosen according to three criteria.  The subset should: contain targets of high 7\m\ surface brightness (to ensure detectable targets in the near-infrared); span a range of properties like metallicity, infrared-to-blue ratio, mid-infrared color, morphology, etc.; and be constrained in redshift such that multiple galaxies could be observed with the same grating tilts, to minimize observational overheads.

\subsection{IC~10}
IC~10 is classified as a Magellanic-type dwarf irregular galaxy
and has been extensively studied in the optical (e.g. Lequeux et al. 1994; Hunter 2001).  The galaxy appears to exhibit ongoing star formation at a rate of $0.15~M_\odot~{\rm yr}^{-1}$ (Thronson et al. 1990) and has the highest global surface density of Wolf-Rayet stars in the Local Group (Massey \& Armandroff 1995).  The \HI\ emission spans some seven times the optical extent of the galaxy, and Wilcots \& Miller (1998) suggest that IC~10 is still forming through accretion of the surrounding material.

\subsection{NGC~0693}
NGC~693 is a non-Magellanic irregular system with average infrared-to-optical and infrared colors.  This galaxy's fairly high global \colore\ ratio is presumably indicative of a system dominated by diffuse cirrus emission (a low \colore\ ratio would indicate intense star formation; Dale et al. 2000).  This impression is supported by the
optical spectrum (integrated along the major axis), which shows strong stellar absorption features in the blue (e.g. calcium H and K, and higher-order Balmer lines).  

\subsection{UGC~2855}
UGC~2855 is a late-type spiral with prominent arms that are quite striking in the mid-infrared  (Dale et al. 2000).  H\"{u}ttemeister, Aalto, \& Wall (1999) argue that this galaxy  lacks evidence for strong shocks and is in a pre-burst stage.  Their claim is based on the $\sim 8$~kpc-long continuous molecular bar, which is relatively undisturbed and rotating in a solid-body fashion.  Moreover, the \COa/\COc\ ratio is low and the \hal\ emission along the bar is weak.  Our optical spectrum of a representative \ion{H}{2} region in UGC~2855 shows very strong nebular emission lines superposed on a weak and relatively featureless underlying continuum.  UGC~2866 is a companion galaxy.

\subsection{NGC~1569}
The post-starburst NGC~1569 is famous for its two brightest compact super-star clusters (``A'' and ``B''). In this study we have targeted two of the galaxy's \HII\ regions.  The southeastern and northwestern \HII\ regions respectively correspond to regions 7 and 2 of Waller (1991); the latter has also been designated as ``Region~C'' by Greve et al. (1996) and ``GMC~3'' by Taylor et al. (1999).  NGC~1569 has unusually strong ionic line intensities relative to measures of the dust and photodissociation region emission, in both the mid- and far-infrared.  It stands out as extreme among the general Key Project sample in the ratios of [\ion{O}{3}]~88\m/far-infrared, [\ion{O}{3}]~88\m/[\ion{C}{2}]~158\m, and [\ion{O}{3}]~88\m/[\ion{O}{1}]~63\m, consistent with the idea that the galaxy is powered by vigorous ongoing star formation (Malhotra et al. 2001).  The extreme mid-infrared colors point to an intense radiation field (Dale et al. 2000).  The $2.5-11.6$\m\ ISOPHOT spectra from 24\arcsec$\times$24\arcsec\ apertures centered on the southeastern and northwestern \HII\ regions (but overlapping with the super-star clusters) are dominated by extraordinarily strong ionic emission, in this case the [\ion{S}{4}]~10.5\m\ line, relative to the underlying dust continuum emission.  The optical spectra of \HII\ regions in this galaxy display extremely strong [\ion{O}{3}]~4959,~5007~\AA\ emission due to the high electron temperature $T_{\rm e}$, a result of the low abundance of oxygen, the primary coolant.  However, since the infrared lines (see above) are not sensitive to $T_{\rm e}$, the strong line-to-continuum contrast contrast in the infrared must have a different origin, possibly a deficiency of interstellar dust emitting at these wavelengths due to the extremely intense ultraviolet radiation field (see Lu et al. 2003).





\subsection{NGC~2388}
NGC~2388 is a spiral galaxy with fairly unremarkable infrared properties.  We did not obtain an optical spectrum of NGC~2388.

\subsection{NGC~4418}
\label{sec:n4418}
NGC~4418 has the most unusual infrared properties in our sample.  Roche et al. (1986) interpret the deep 9.7\m\ spectral trough in its spectrum as a silicate absorption feature, and infer a high extinction, $A_V>50$, towards the nuclear source.  The galaxy has an extremely small \colore\ ratio and an extremely large \IRAScolor\ ratio, indicative of an intense interstellar radiation field (Dale et al. 2000).  Its mid-infrared spectrum lacks the typical polycyclic aromatic hydrocarbon features (Lu et al. 2003).  NGC~4418 also exhibits an unusually low \CII/far-infrared ratio (an upper limit, in fact); the interpretation favored by Malhotra et al. (2001) is that interstellar dust grains are highly positively charged in this active galaxy, leading to less efficient heating by photo-ejected electrons from dust grains.  The optical spectrum of NGC~4418 is dominated by spectral features from late-type stars.  The only emission line evident is a weak emission peak near 6620~\AA\ which may be a component of
a complete H$\alpha$ profile that includes a small absorption dip at the galaxy's systemic velocity.  Due to the large extinction towards the infrared-active nuclear region of this galaxy, the optical spectrum does not probe this region, but instead simply presents a spectrum characteristic of the older stellar population of the galaxy's disk.
                                                                                

\subsection{NGC~6946}
NGC~6946 is a well-studied, nearly face-on galaxy with classic spiral structure and a starbursting core (Engelbracht et al. 1996).  Schlegel, Blair, \& Fesen (2000) find 14 pointlike X-ray sources in NGC~6946, the brightest of which is unremarkable in the mid-infrared (compare with the mid-infrared maps in Dale et al. 1999).  The \HII\ region we targeted to the NE shows up in the $ROSAT$ HRI image as a faint arc $\sim$0\farcm5 in length, whereas the nucleus is associated with a point source with the third highest X-ray luminosity in the galaxy ($L_X(0.5-2.0~{\rm keV})\sim29\cdot10^{30}$~J).

\subsection{NGC~7218}
NGC~7218 is a late-type spiral with typical infrared properties.  The galaxy has a short bar oriented roughly perpendicular to the major axis.  The optical emission lines show a level of star-forming activity intermediate between NGC~693 and NGC~7771 on the one hand, and UGC~2855 and NGC~1569 on the other.
                                                                                
\subsection{NGC~7771}
\label{sec:n7771}
NGC~7771 has a prominent circumnuclear star-forming ring of major axis diameter 6-7\arcsec\ with $\approx$10 radio-bright clumps and $\approx$10 near-infrared-bright clumps (Smith et al. 1999; Reunanen et al. 2000).  The radio and near-infrared sources in the ring are offset from each other, lending to the idea that there are multiple star-forming populations of different ages.  A possible interaction with the nearby dwarf NGC~7770 may be related to much of this activity.  Smith et al. (1999) find that the radio regions are $\approx$4~Myr old, whereas the near-infrared spots are slightly older, some $\approx$10~Myr.  Reunanen et al. (2000) present \Brg\ and \Htwo\ line maps and find that the starburst ring exhibits small \Brg\ equivalent widths 
(of order unity dex) and their interpretation is an instantaneous starburst that occurred 6-7~Myr ago.  

\section{OBSERVATIONS AND DATA REDUCTION}

\subsection{Near-Infrared Spectroscopy}
The spectroscopic observations were carried out between September~2000 and November~2001 (Table~\ref{tab:obs}) using the Palomar Integral Field Spectrograph (PIFS; Murphy, Matthews, \& Soifer 1999) on the Palomar 200-inch Telescope.  PIFS provides a 5\farcs4$\times$9\farcs6 field of view using eight separate 0\farcs67$\times$9\farcs6 slits to feed two independent spectrographs within the same liquid N$_2$-cooled dewar.  Each slit is four pixels wide and 58 pixels long (0\farcs167~pixel$^{-1}$).  The high resolution mode ($R\sim1000-1500$) was used, providing a usable wavelength range of 0.03--0.04\m\ and a velocity resolution of $\Delta v \sim 200-300$\kms\ near the observed wavelengths.

The standard observational procedure consisted of a set of four five-minute on-off integrations.  Thus 20~minutes were spent integrating on source, and an equal amount of time off source.  The only exceptions to this procedure involved UGC~2855 \Htwo\ and NGC~693 \Brg, for which we observed 40~minutes and 45~minutes on source, respectively.  For all observations, the 5\farcs4$\times$9\farcs6 field of view was oriented with the long axis at a position angle of 90\arcdeg.  A positional dithering pattern was employed for the sequence of observations to recover seeing-limited spatial sampling in the cross-slit direction.  This dithering is the reason the resultant maps (5\farcs7$\times$10\farcs0) are slightly larger than the PIFS field of view.  Spectral calibration lamp data were taken immediately after each set of four dithered integrations.
Corrections for atmospheric opacity and spectral flat-fielding are derived from observations of F8-G9 main sequence stars.  An estimate of the near-infrared PSF was periodically obtained, with each measurement comprised of four dithered exposures of a nearby field star using the PIFS imaging mode (see Table~\ref{tab:obs}).

The atmospheric conditions for each set of observations are listed in Table~\ref{tab:obs}.  Since about half of the observations were taken in non-photometric conditions, for uniformity all the observations are
flux-calibrated using 2MASS $JHK$ data, via fluxes from cutouts matched to the PIFS field of view.  This technique is uncertain.  First, near-infrared broadband fluxes are contaminated at some level by nebular line emission.  In very young starbursts/blue compact galaxies, for example, the line contribution can be as large as $\sim$5-10\% of the total near-infrared flux (e.g. Figure~3 of Kr\"{u}ger, Fritze-v. Alvensleben, \& Loose 1995; Vanzi et al. 2000).  However, except for IC~10 \Pab, NGC~1569~NW \Pab, and NGC~6946~NE \Brg, none of the equivalent widths for our sample indicate broadband flux contributions larger than 2\%.  Second, the 2MASS and PIFS aperture positionings may not exactly coincide.  Third, depending on the location of the emission within the PIFS field of view, 2MASS flux dilution can be a factor since the 2MASS images are $\sim3$\arcsec\ in resolution whereas the PIFS data typically are sub-arcsecond.  Based on a comparison of 2MASS flux calibration and the calibration we obtained from standard stars on the photometric nights, we estimate near-infrared fluxes in this work are uncertain at the $\sim50$\% level.  Figures~\ref{fig:pifs1}--\ref{fig:pifs4} present line emission contour maps superposed on the continuum images at the same wavelength/grating setting, whereas Figures~\ref{fig:spectra1} and \ref{fig:spectra2} display the integrated spectra.  A few of the spectral images exhibit strong residuals due to non-uniform illumination during the spectral flat-field procedure.

\subsection{\hal\ Imaging}
\label{sec:Halpha}
\hal\ images of the galaxies in this sample were obtained in several observing runs.  NGC~2388 was observed with the 1.07~m Hall telescope at Lowell Observatory, and the rest of the sample was observed with the Perkins 1.8~m telescope at Lowell Observatory.  IC~10 and NGC~1569 were observed in November 1992 with a TI 800$\times$800 CCD on loan to Lowell Observatory from the US Naval Observatory in conjunction with the Ohio State University Fabry-Perot that was used simply as a 3:1 focal reducer. The resulting scale was 0\farcs488~pixel$^{-1}$.  NGC~2388 was observed in December 1997 with a SITe 2048$\times$2048 CCD, binned 4$\times$4.  The pixel scale was 0\farcs62~pixel$^{-1}$.  The rest of the observations with the Perkins telescope were made in October--December 1996 and used an 800$\times$800 TI CCD provided to Lowell Observatory by the National Science Foundation.  This detector was coupled to a 4:1 focal reducer to yield a pixel scale of 0\farcs433~pixel$^{-1}$.  NGC~6946 was observed with the KPNO 0.9~m telescope in May~1993, and the data were graciously provided by A. Ferguson (Ferguson et al. 1998).  The continuum was subtracted using an image taken with a narrowband filter centered at 6409~\AA\ and width 88~\AA\ (A. Ferguson, private communication).  January~200 KPNO 2.1~m \hal\ data of NGC~1569 were kindly provided by C. Martin to enhance our analysis (Martin, Kobulnicky, \& Heckman 2002).

All observations at Lowell Observatory used a series of redshifted \hal\ filters with FWHM of 27--32~\AA, and an off-band filter centered at 6440~\AA\ with a FWHM of 95~\AA.  The off-band image was shifted, scaled, and subtracted from the \hal\ image to remove the stellar continuum and leave only \hal\ nebular emission.  The images were reduced using the over-scan strip to remove the electronic pedestal and dome flat-field images to remove pixel-to-pixel variations.  Multiple images were obtained and combined to remove cosmic rays.

The \hal\ emission was calibrated using the known \hal\ flux of NGC~604 (D. Hunter, unpublished spectrophotometry) and spectrophotometric standard stars with minimal \hal\ absorption features (Stone 1977; Oke \& Gunn 1983).  These objects agreed with each other to 4\% and the calibrations remained constant to 8\% for a given filter over many observing runs.  The \hal\ flux has been corrected for \NII\ contamination and the position of the line within the bandpass, shifted for the outside temperature.  We assume \hal/(\hal\ + \NII)$\approx0.75$ (Kennicutt 1992; Jansen et al. 2000).  To account for the temperature-dependent shifting of the filter bandpass, we assume the bandpass shifts blueward by 0.18~\AA\ for every degree Celsius colder than 20\degr~C. Fluxes for the PIFS fields-of-view are extracted from the \hal\ images.  The overall absolute calibration of these fluxes is estimated at $\sim30$\% to account for PIFS--\hal\ aperture misalignment, unknown observing systematics, and the uncertainties in \NII\ contamination, which exhibits a 20\% dispersion in star-forming galaxies (Kennicutt \& Kent 1983; Kennicutt 1992; Jansen et al. 2000).

\subsection{Optical Spectroscopy and Derived Abundances}
\label{sec:optical_spectroscopy}

Optical long-slit spectra of six of the galaxies studied in this paper,
NGC~693, NGC~1569, NGC~4418, NGC~7218, NGC~7771, and UGC~2855, were obtained with the 
Large Cassegrain Spectrograph (LCS) on the McDonald Observatory Harlan J. 
Smith reflector during observing runs in September~1996, March~1997,
and September~1997.  These data are a subset of a larger body of 
optical spectra of galaxies in the {\it ISO} Key Project, which will be
presented and discussed in more detail in a future paper (Dinerstein 
et al., in preparation).  
The 90$\arcsec$-long slit of the LCS was opened to a  
width of 1\farcs5, yielding a spectral resolving power of  
$\sim$800 on the Texas Instruments 800$\times$800-pixel CCD detector.
Separate setups were used to cover the blue ($\sim$~3680$-$5060~\AA)
and red ($\sim$~4800$-$7450~\AA) spectral regions; both set-ups
included H$\beta$ and the [\ion{O}{3}] 4959, 5007 \AA\ lines. 
Both blue and red data were obtained for all of the galaxies except 
NGC~1569 and UGC~2855, for which only the red spectra were taken.
Integration times are given in Table~\ref{tab:optical_spectroscopy}.
Standard IRAF procedures were used to reduce the data, extract and calibrate
apertures corresponding to selected spatial regions, and
measure line strengths.

The physical regions sampled by the optical spectra vary 
from galaxy to galaxy, and do not always correspond to
the apertures used for the near-infrared spectroscopy.
The optical slit was usually rotated to the position
angle of the galaxy's major axis, but in some cases 
we then offset onto emission peaks in the \hal\ images described in \S~\ref{sec:Halpha}. 
For NGC~693, NGC~4418, and NGC~7771, a single spectrum 
representing the integrated light of the disk was
extracted, while for NGC~1569, NGC~7218, and UGC~2855
we extracted spectra from smaller apertures centered on bright disk \ion{H}{2} regions.

The primary quantities we derived from the optical  
spectra are tabulated in Table~\ref{tab:optical_spectroscopy}: $c$, the logarithmic
extinction at H$\beta$ as derived from the ratio of
H$\alpha$ to H$\beta$ (which we used to correct the optical
line ratios for reddening); the measured H$\beta$
flux in the extraction aperture (uncorrected for 
extinction); 
and several bright-line 
indices used as empirical metallicity indicators.  These include:
R$_{23}$~=~([\ion{O}{3}]~4959,~5007~$+$~[\ion{O}{2}]~3727)/H$\beta$;
R$_{3}$~=~([\ion{O}{3}]~4959,~5007)/H$\beta$; 
O$_{32}$~=~([\ion{O}{3}]~4959,~5007)/[\ion{O}{2}]~3727; and
N/O~=~[\ion{N}{2}]~6584/[\ion{O}{2}]~3727.
The [\ion{S}{2}]~6717,~6731~\AA\ intensity ratios generally indicate low values for the
electron
density $n_{\rm e}$, $\log~n_e\sim2.0\pm0.3~({\rm cm}^{-3}$).
  
For the three galaxies for which we obtained blue data,
NGC~693, NGC~7218, and NGC~7771, 
we used the R$_{23}$ calibration  
of Edmunds \& Pagel (1984) 
to obtain the (O/H) abundances given in Table~\ref{tab:sample}.
If we had used the calibrations of 
McGaugh (1991, 1994)
for intermediate ionization parameters, 
the inferred (O/H) values would be systematically 
lower by $-$0.1~dex, but the galaxies would retain the  
same order and differences in abundance values. 
Note the contrast 
between NGC~693 and NGC~7771, which have similar line
fluxes and line-to-continuum contrast (in integrated light)
but strikingly different ionic line ratios (Table~\ref{tab:optical_spectroscopy}),
which indicate that (O/H) and (N/O) are  
about 0.4$-$0.5~dex higher in NGC~7771 than in NGC~693.
These two galaxies give an excellent illlustration
of both the excitation-metallicity
effect which is the basis of the bright-line method,
and the well-known correlation between metallicity
and total mass or luminosity of disk galaxies
(see Garnett 2002 for a recent discussion).  NGC~7218 
is intermediate in abundance between NGC~693 and NGC~7771.
Since we have only red spectra for UGC~2855, 
we assume that it has the same (O/H) as NGC~693 
based on the similarity of their R$_3$ indices. 
On the other hand, NGC~1569 has extremely strong
[\ion{O}{3}] lines. Taking R$_3$ as a lower
limit to R$_{23}$, we find an (O/H) value of about 8.2, consistent
with the result of Kobulnicky \& Skillman (1997). 

\section{RESULTS}

\subsection{Fluxes, Equivalent Widths, and Starburst Ages}
\label{sec:fluxes}
Near-infrared emission lines are securely detected in 84\% (37/45) of the lines and regions targeted (Table~\ref{tab:fluxes}).  
In general, \Pab\ is the strongest of the observed lines, and \Htwo~(2.122\m) the weakest (no significant \Htwo\ emission is observed for any \HII\ region in the sample).  
This is also true for the extinction-corrected fluxes (\S~\ref{sec:extinction}).  The most notable exception to this rule is provided by NGC~4418, which has a surprisingly large \Htwo\ flux, brighter in fact than the other three near-infrared lines observed combined.

The equivalent widths are generally smaller for the nuclei than for the \HII\ regions.  For example, the uncertainty-weighted averages for \Pab\ and \Brg\ are $2.9\pm1.5$~\AA\ and $5.3\pm0.9$~\AA\ for nuclei and $40.\pm15$~\AA\ and $24\pm6$~\AA\ for \HII\ regions.  This is no surprise since the nuclear regions have comparatively much stronger near-infrared continuum emission from older stars.  Ho, Fillipenko, \& Sargent (1997) found a similar result in their study of \HII\ nuclei and disk \HII\ regions. 

The \Pab\ and \Brg\ equivalent widths are essentially ratios of line emission from \HII\ regions around young stars to continuum emission from older stars.  If the conditions of either an instantaneous starburst or recent constant star formation are met, then these equivalent widths are excellent age indicators.  From a modeler's perspective, one needs to know the star formation history for at least 1~Gyr; if there are variations in the star formation rate with time, the continuum is difficult to constrain.  On the other hand, if the recent starburst is dominant and has an age less than 30~Myr, its red supergiant population is often so strong that the older population does not significantly affect the inferred starburst age.  For example, in the $10^6~M_\odot$ instantaneous starburst model of Starburst99 (Figure~51 of Leitherer et al. 1999), the peak in the $K$ band luminosity as a function of time (for $Z\gtrsim0.4Z_\odot$) stems from the burst's red supergiant population; this peak at $10^7$~yr characterizes the starburst.  Compare this luminosity to the luminosity generated by a galaxy-wide continuous $1~M_\odot$~yr$^{-1}$ star formation that occurs for, say, $10^9$~yr (see their Figure~52).  Considering that only a small portion of this underlying global luminosity is projected to spatially overlap with an individual \HII\ region, \Brg\ equivalent widths for an \HII\ region can indeed be relatively unaffected by an older population.

Assuming that the \Brg\ equivalent width can be related to the effective star formation age, the ages derived for an instantaneous starburst (Leitherer et al. 1999) range from a few million years for the youngest \HII\ regions to longer than 10~Myr for the oldest nuclear regions (Table~\ref{tab:results}).  The \HII\ regions tend to have more recent starbursts than nuclei.  If nuclei in general are not defined by a single starbursting population, and instead are comprised of star formation episodes that span a range of ages, then it is expected that their inferred ages would be older than an isolated starburst.  Interestingly, excluding the unusual system NGC~4418, there is perhaps a trend in the starburst age with mid-infrared color.  Systems with lower \colore, indicating more active star formation, tend to be younger (see Figure~\ref{fig:density}).  A similar trend was previously reported for a study of galaxy circumnuclear regions (Roussel et al. 2001).  
Finally, readers familiar with post-starburst NGC~1569 may be surprised by its apparently young age.  We emphasize that our observations centered on two \HII\ regions, not the famous super-star clusters.  Moreover, the differing spatial distributions in the continuum and line (\S~\ref{sec:spatial_distribution}) can lead to ill-defined equivalent widths; we have extracted an average equivalent width, based on the entire fluxes within the PIFS field of view.

\subsection{Spatial Distribution}
\label{sec:spatial_distribution}
For the nuclear regions where line emission is detected, the continuum and line emission are generally co-spatial (have similar morphology).  On the other hand, there are clear spatial offsets between the line emission from the ionized gas and the continuum emission in the disk \HII\ regions (see the maps in Figure~\ref{fig:pifs1}--\ref{fig:pifs4}).  

The similar distributions of the line (gas) and continuum emission (stars) in the galaxy nuclei may arise naturally, as a result of interstellar material settling to the center of the galaxy's gravitational potential well, confining the peak density of various components to a narrow region.  Conversely, the \HII\ regions do not provide the same kind of gravitational environment, so that the distribution of ionized gas is affected by a number of other factors.  For example, interstellar material can be blown away from a region that is actively forming stars: the \hal\ map of NGC~1569 exhibits a hole centered on the two bright super starclusters, and the distinct ``S'' shape to the near-infrared line emission maps in NGC~1569~NW is also apparent in \hal\ at the same location (Martin, Kobulnicky, \& Heckman 2002; the emission line contours for NGC~1569~NW correspond to GMC~3 of Taylor et al. 1999).  

Alternatively, there may simply be two or more populations of different ages, as in the Smith et al. (1999) study of the NGC~7771 circumnuclear star-forming ring and the Alonso-Herrero et al. (2003) study of starbursts M~82 and NGC~253.  In NGC~7771, Smith et al. (1999) find that the 1.6~kpc diameter ring appears as $\approx$10 distinct sources in both the radio and near-infrared continua.  However, the radio and near-infrared clumps do not overlap, suggesting at least two different formation timescales.  Most of the ring is visible in the PIFS field of view, and there is evidence for at least three distinct sources in our \Pab\ map.  Alonso-Herrero et al. (2003) find a continuum-line spatial mismatch for much of M~82 and NGC~253 in Pa$\alpha$ and \FeIIb.  In addition, given that the majority of radio supernova remnants and compact \FeIIb\ sources are not spatially aligned, these authors suggest that there is a young and old population of supernova remnants.

\subsection{Extinction}
\label{sec:extinction}
We derived the extinction for each region from the ratios of \hal\ (from the narrow-band imaging), \Pab, and \Brg.  The values listed in Table~\ref{tab:results} are derived by least-squares fitting a screen model to the observed line ratios \hal/\Brg\ and \Pab/\Brg\ (the two line ratios are fit simultaneously) and assuming the Li \& Draine (2001) extinction curve.  Uncertainties on the derived extinctions are based on the observational uncertainties and the predicted range in the intrinsic line ratios: \hal/\Brg$=97.5\pm5.5$ and \Pab/\Brg$=5.75\pm0.15$ for Case B recombination; $100<n_{\rm e}({\rm cm}^{-3})<10,000$; and $5,000<T_{\rm e}({\rm K})<10,000$ (Storey \& Hummer 1995).

A small difference is found between the nuclear and \HII\ region extinctions.  The averages for the sample are $A_V\approx1.4\pm$0.7~mag for the \HII\ regions and 2.7$\pm$0.6~mag for the nuclei.  Our results are consistent with previous determinations.  
For example, Reunanen et al. (2000) find $A_V$(int+MW)$=2.8$~mag for the central 11\arcsec$\times$11\arcsec\ of NGC~7771, and Davies, Alonso-Herrero, \& Ward (1997) obtain $A_V=3.0$~mag for the more obscured optical nucleus (their ``region 2'').  Engelbracht et al. (1996) find that the core of NGC~6946 is extincted by $A_V$(int+MW)$=4.3$~mag for a foreground screen model.
The internal extinction for NGC~1569~SE and NGC~1569~NW has been estimated by Kobulnicky \& Skillman (1997) to be $\sim$0.2~mag (their regions C15 and A12, respectively).  The extinction for NGC~4418 is unconstrained by these data, since the fluxes for both \Pab\ and \Brg\ are upper limits.  

The range of extinctions found by Ho, Filippenko, \& Sargent (1997) for a sample of over 200 ``\HII\ nuclei'' is $0 \lesssim A_V \lesssim 3$.  The results of Kennicutt, Keel, \& Blaha (1989) for \HII\ and starburst nuclei are consistent with this result.  This range of extinction for optically-selected samples is a bit smaller than the average nuclear extinction derived for our sample ($\left< A_V (\rm {int}) \right>\sim $3~mag), with NGC~693 being significantly higher than the average.  In fact, in most cases the extinction values are larger than those indicated by our long-slit-derived ratio of H$\beta$/H$\alpha$ (note: for a ratio of total to selective extinction of $R=A_V/E(B-V)=3.1$, $A_V=2.1~c$).  This is to be expected in regions where dust is mixed with gas, since the optical line emission is seen selectively from the regions of lowest extinction.  As Kennicutt (1998) points out, infrared-selected samples reveal much more dramatic nuclear star formation, and $V$ band obscuration can easily exceed several magnitudes in the cores of star-forming galaxies; we are able to probe deeper at near-infrared wavelengths than optical studies, and the more deeply embedded sources simply are more obscured.

\subsection{Star Formation and The Interstellar Radiation Field}
The number of hydrogen-ionizing photons is derived from an average of the (extinction-corrected) \Brg\ and \Pab\ luminosities
\be
N({\rm H^o}) [s^{-1}]=\onehalf \{ 7.63 L_{\rm Br\gamma} [10^{20}~W] + 1.40 L_{\rm Pa\beta} [10^{20}~W] \}
\ee
(Leitherer \& Heckman 1995).  If either the \Brg\ or \Pab\ flux is an upper limit, then only the securely detected flux is used.  Table~\ref{tab:results} also expresses this number in terms of the equivalent number of O8.5~V stars, using a conversion from Vacca, Garmany, \& Shull (1996).  The average projected density of ionizing photons for this sample, computed from a ratio of the number of ionizing photons to the size of the emitting region, spans two orders of magnitude, from $\log N({\rm H^o}) [{\rm s}^{-1}~{\rm pc}^{-2}] =46.4$ for NGC~7218 to 48.2 for NGC~1569~NW (the uncertainty on these numbers is 0.2 dex).  A strong and understandable correlation is obtained between the mid-infrared color \colore\ and ionizing photon density: lower \colore, an indicator of more active star formation, corresponds to more intense radiation environments (Figure~\ref{fig:density}).  F\"{o}rster Schreiber et al. (in preparation) find a similar result for disk and nuclear regions in a sample of nearby galaxies.  Perhaps not surprisingly, the one outlier in the observed trend is NGC~4418, the most unusual target in our sample (see \S~\ref{sec:n4418}, \ref{sec:fluxes}, and \ref{sec:extinction}).  This object may falsely appear to be lacking in ionizing photons either because of extreme extinction, or because its \HII\ regions are very young and still rich in dust which absorbs a large fraction of the ionizing photons.  Finally, with the exception of the \HII\ region dubbed ``NGC~6946~NE,'' the \HII\ regions exhibit the 
most intense radiation fields and lowest \colore\ ratios.  

The combination of $N({\rm H^o})$ and the \Htwo\ flux can place constraints on the dominant excitation mechanism for \Htwo.  Based on the results of Black \& van Dishoeck (1987), $I_{\rm UV}/I_{\rm H_2}\sim10^5-10^8$ for \Htwo\ fluorescence triggered by ultraviolet radiation (\Htwo\ fluorescence, however, can initially be much more significant compared to its final equilibrium value, e.g. Goldshmidt \& Sternberg 1995 and Hollenbach \& Natta 1995).  For our sample, the observed ratio is of order $10^4$, suggesting that the dominant excitation mechanism is shocks rather than fluorescence.

\subsection{Near-Infrared Line Ratio Diagnostics}
The \FeII/\Pab\ line ratio can be a useful diagnostic of the star formation history since \FeII\ emission is thought to arise from supernova activity.  Accordingly, \FeII\ is an indicator of past star formation, whereas \Pab\ signifies current star formation (e.g. Moorwood \& Oliva 1988; Calzetti 1997).  In a sense \FeII/\Pab\ is a ratio of non-thermal to thermal activity (however, both thermal and non-thermal mechanisms can be attributed to supernovae and their remnants, e.g. Sankrit et al. 1998; Rho et al. 2003).  While other lines also trace the current star formation, \Pab\ is particularly useful since \FeIIa/\Pab\ is relatively insensitive to extinction.  Infrared-bright galaxies that are predominantly powered by star formation should thus have lower \FeII/\Pab\ ratios than AGN-dominated systems (e.g. Larkin et al. 1998).

\Htwo(2.122$\micron$)-to-\Brg\ is another ratio that in principle is unbiased by extinction.  Its interpretation, however, is more difficult.  \Htwo, in contrast to \FeII, can be triggered by both thermal and non-thermal processes.  As described by Calzetti (1997), molecular hydrogen on the surface of molecular clouds can be collisionally excited in gas heated by slow shocks, ultraviolet radiation, and X-ray radiation, or it can result from fluorescence following ultraviolet absorption.  Empirically, starburst galaxies and \HII\ regions exhibit lower \Htwo/\Brg\ ratios (Larkin et al. 1998).  

Inspection of Figure~\ref{fig:ratios} confirms that these line ratios help to empirically distinguish between galaxies powered by different processes.  Data from the literature show there is an approximate trend proceeding from the upper right to the lower left of increasing star formation activity and decreasing AGN activity (or, in the absence of AGN activity, decreasing star formation activity occurring in brief, intense and embedded bursts with shocks).  In fact, all three \HII\ regions for which we measured both line ratios reside in the lower left, and the one source in our sample suggested to harbor AGN activity, NGC~4418 (Roche et al. 1986; Baan, Salzer, \& LeWinter 1998; Spoon et al. 2001), is the upper-rightmost data point in our sample.  The nuclear observations for the remainder of our sample lie between these extremes.  It should be noted, however, that systems with low metallicity might be expected to have low iron-to-hydrogen ratios regardless of the source that powers their infrared emission; the location of IC~10 and NGC~1569 in Figure~\ref{fig:ratios} must reflect in part their low metallicity.

\subsection{Singly-Ionized Iron-to-Hydrogen Mass Ratio}
Following F\"{o}rster Schreiber et al. (2001), the gas-phase abundance of Fe$^{1+}$ can be estimated assuming a single-layer model of uniform density and temperature:
\be
{n({\rm Fe^{1+}}) \over n({\rm H})} = {L([{\rm FeII}]) j({\rm H}) \over L({\rm H}) j([{\rm FeII}])}
\label{eq:abundance}
\ee
where $L$ is the observed line luminosity, $n$ the number density, and $j$ the line emissivity.  \Brg\ is the reference H line in this work.  The emissivities are computed in the low density limit\footnote{The critical densities of the observed Fe transitions are of order $10^4$~cm$^{-3}$ (Hamann \& Ferland 1999; Quinet, Le Dourneuf, \& Zeippen 1996), so we assume collisional de-excitation is unimportant when computing line emissivities.} assuming $T_{\rm e}=5000$~K (Storey \& Hummer 1995; Hamann \& Ferland 1999).  The Fe$^{1+}$ abundances are listed in Table~\ref{tab:results}.  For comparison, the total gas-phase iron abundance in the Orion Nebula is estimated to be of order Fe/H$\sim 10^{-6}$ (see Rubin et al. 1997 and references therein).

\subsection{NGC~4418}

One particular object exhibits extreme properties in essentially every category.  NGC~4418 is isolated in plots of near-infrared line ratios, ionizing photon density and starburst age.  Unlike every other source, it was undetected in \Pab\ and \Brg, and \Htwo\ is the brightest of the four transitions sampled.  It is possible that NGC~4418 deviates from the overall age-infrared color trend due to a more complicated star formation history.  Some claim the galaxy harbors a heavily obscured AGN (Roche et al. 1986; Spoon et al. 2001), while others suggest it is a pre-starburst galaxy.  Roussel et al. (2003), for example, use new multi-wavelength radio data to argue that galaxies like NGC~4418 may be experiencing the very onset of a starburst after a long time of quiescence, giving them a distinctive infrared emission signature in both continuum and lines.  Other studies have shown NGC~4418 to likewise be peculiar (\S~\ref{sec:n4418}).  Though our data do not have the spectral and spatial resolution to resolve this debate, they similarly point to an object that is not currently powered by star formation.  


\section{SUMMARY AND DISCUSSION}
Near-infrared spectroscopy over two spatial dimensions has been obtained for the first time for nuclear and extranuclear star-forming regions in a diverse sample of nine nearby galaxies.  The sample was previously observed with $ISO$ as part of the Key Project on the Interstellar Medium of Normal Galaxies.  Mid-infrared and other ancillary data are used in conjunction with continuum and line maps at \Pab, \Brg, \FeII, and \Htwo\ wavelengths to explore the physical processes underlying star formation.  

In this small sample, the \HII\ regions differ from the galaxy nuclei according to nearly every metric.  Nuclear observations typically show spatial coincidence of the continuum and line emission, unlike that seen for star-forming extranuclear regions.  In addition, the star-forming episodes characterizing the \HII\ regions appear to be younger and more intense than in the nuclei.  One hypothesis is that the density of material in nuclear regions supplies continual, on-going episodes of star formation, while extranuclear star-forming regions are more transient and dependent upon random overdensities in the interstellar distribution.  Thus if measures of a single star-formation age are made, such as that inferred from \Brg\ equivalent widths, on-going nuclear star formation would naturally lead to longer timescales than that seen for individual starbursting episodes in \HII\ regions.  Moreover, continual nuclear star formation means both the near-infrared continuum and line transitions would be strong at the same locations, even if they represent stellar populations of different ages.  Alternatively, though there may be cycling in the nuclear star formation history, the strong central density gradient should result in spatially coincident continuum and line emission.  Though the \HII\ regions appear to differ from the nuclei by most measures, this result is perhaps not surprising nor profound, especially given our limited sample of targets.

Near-infrared line ratios that distinguish between AGN and star-forming activity are consistent with the \HII\ regions being dominated by star formation whereas the nuclear regions reflect a mixture of star formation and a secondary ionization component (in the absence of AGN activity, the secondary component could be due to embedded and intense starburst-induced shocks).  Perhaps the nuclei are not exclusively powered by normal star formation, echoing the suggestion of Terlevich et al. (1992) that AGN activity is the natural consequence of the final stages of intense nuclear star formation.  Ho, Filippenko, \& Sargent (1997) find that about half (42\%) of the bright spirals have nuclear emission spectra with \HII\ region-like line ratios; the percentage is a steep function of global morphology, with up to 80\% of the later types exhibiting \HII\ nuclei.  Kennicutt, Keel, \& Blaha (1989) find that approximately ``half of the \HII\ region and starburst nuclei show evidence for a secondary ionization component, either an active galactic nucleus or large-scale shocks ... [supporting] the idea that many spiral nuclei are composite in nature, with a central LINER or Seyfert-like nucleus surrounded by star-forming regions.''  Consistent with this interpretation, our nuclear data do not exhibit extreme line ratios indicative of pure star formation or AGN dominance, but instead suggest a combination of the two processes.

One of the secondary goals of this work was to search for any trends in the sample as a function of metallicity.  No obvious correlations are seen with respect to the oxygen abundances.  However, on a purely empirical basis the \HII\ regions show lower gas-phase Fe$^{1+}$ abundances than nuclei by an order of magnitude.  {\it Total} gas-phase Fe abundances cannot be computed with these data since only the singly-ionized species has been observed.  These data are additionally limited since Fe$^{1+}$ abundances are difficult to interpret.  Not only do Fe$^{1+}$ abundances depend on the overall amount of Fe produced, Fe$^{1+}$ abundances also depend on the ionizing strength of the interstellar radiation field 
and how much Fe has been depleted onto, or released from, dust grains.  These complications are likely why the abundances do not correspond in any way to the derived metallicities.  Nevertheless, there is an interesting contrast in the Fe$^{1+}$/H abundance for the \HII-like regions and the nuclei.  The \HII-like regions show low abundances in comparison to the nuclei.  
Lack of abundance data from additional ionic species leaves many questions unanswered.
For example, do environments with harder radiation fields like \HII\ regions exhibit relatively higher Fe$^{2+}$ and Fe$^{3+}$ abundances, helping to explain their relatively low Fe$^{1+}$ abundances?  Models of Orion Nebula data indirectly support this scenario, with
ionization fractions of Fe$^{1+}$/Fe$^{2+}$/Fe$^{3+}$=0.0128/0.244/0.744 (Baldwin et al. 1991) and 0.0529/0.414/0.533 (Rubin et al. 1991).  To directly compare these fractions in a relatively soft and a relatively hard radiation environment, we utilize the NEBULA software (Rubin et al. 1991 as updated recently Rodr\'\i guez \& Rubin 2003).  Keeping the metallicity (Orion-like) and the number of ionizing photons (10$^{50}~{\rm s}^{-1}$) the same from simulation to simulation, we find 
a relatively small change in the Fe$^{1+}$ abundance but a clear shift to higher Fe$^{3+}$ abundances, mostly at the expense of Fe$^{2+}$, for harder radiation fields (see Table~\ref{tab:ratios}).
                                                                                
Perhaps the most interesting result is that portrayed in the bottom panel of Figure~\ref{fig:density}, the relation between ionizing photon density and mid-infrared color.  Relatively intense radiation fields result in lower \colore, likely via two different effects: reduced emission from polycyclic aromatic hydrocarbons and a steepening mid-infrared continuum at 15\m\ (e.g. Dale et al. 1999; Dale et al. 2001; Helou et al. 2001).  Intense radiation fields could result from higher photon densities or harder radiation (Roussel et al. 2001).  To explore this issue, we turn to O$_{32}$, the ratio of \OIII5007$+$4959 to \OII3727 flux.\footnote{The parameter O$_{32}$ is not the best discriminator for radiation hardness for systems with $12+\log({\rm O/H})\gtrsim8.7$, like NGC~7771 and NGC~6946 (see Figure~8 of Kobulnicky, Kennicutt, \& Pizagno 1999).}  Three de-reddened measurements are listed in Table~\ref{tab:optical_spectroscopy}, and four de-reddened values are drawn from elsewhere: logO$_{32}=0.53, -1.04, 0.88, 0.51$ for IC~10, the nucleus of NGC~6946, NGC~1569~NW, and NGC~1569~SE, respectively (Lequeux et al. 1979; Heckman, Crane, \& Balick 1980\footnote{The value for NGC~6946 was derived assuming that the \OIII4959 flux is 1/3 of the quoted \OIII5007 flux.}; C. Kobulnicky, private communication).  Once again, the \HII\ regions behave differently than the nuclei, logO$_{\rm 32,HII}\gtrsim0.5$ and logO$_{\rm 32,nuc}\lesssim-0.5$, with the \HII\ regions exhibiting harder radiation fields.  This O$_{32}$ segregation according to environment implies that mid-infrared colors are in part determined by the hardness of the radiation field.

\acknowledgements 
We owe gratitude to several who unselfishly contributed: J. Larkin for near-infrared line ratio data; A. Ferguson for an \hal\ image of NGC~6946; C. Kobulnicky, C. Martin and B. Buckalew for NGC~1569 \hal\ images; M. Regan for preliminary investigations of CO data; D.G. Hummer and H.W.W. Spoon for their program that extracts intensity ratios of hydrogenic recombination lines; and C. Leitherer and A. Bianchini for enlightening discussions.  We also would like to acknowledge the Palomar staff and B.T. Soifer who helped make the near-infrared observations possible.  Suggestions by the referee improved this work.  This publication makes use of data products from the Two Micron All Sky Survey, which is a joint project of the University of Massachusetts and the Infrared Processing and Analysis Center/California Institute of Technology, funded by the National Aeronautics and Space Administration and the National Science Foundation.  This research has made use of the NASA/IPAC Extragalactic Database which is operated by JPL/Caltech, under contract with NASA.  IRAF is distributed by the National Optical Astronomy Observatories, which are operated by the Association of Universities for Research in Astronomy, Inc., under cooperative agreement with the National Science Foundation.

\scriptsize
\begin{deluxetable}{lcccccccc}
\def\a{\tablenotemark{a}}
\def\b{\tablenotemark{b}}
\def\c{\tablenotemark{c}}
\def\d{\tablenotemark{d}}
\tablenum{1}
\label{tab:sample}
\tablewidth{510pt}
\tablecaption{The Sample}
\tablehead{
\colhead{Galaxy}    & \colhead{Morph.\a} & \colhead{R.A.}   & \colhead{Decl.}   & \colhead{Dist.\a} &
\colhead{c$z_\odot$}& \colhead{$\log {L_{\rm FIR} \over L_{\rm B}}$\a} & 
\colhead{$\log {L_{FIR} \over L_\odot}$} & \colhead{$\log$O/H}
\\
\colhead{}          & \colhead{}          & \colhead{(J2000)}& \colhead{(J2000)} & \colhead{(Mpc)} &
\colhead{(km s$^{-1}$)} & 
\colhead{}       & \colhead{}        & \colhead{$+12$\c}
}
\startdata
IC 10    & IBm?               & 002024.5 & $+$591730 &~0.8 &$-$348& $-$0.48 & ~8.1 & 8.17~\\
NGC 0693 & I0: sp 	      & 015031.0 & $+$060842 &22.1 & 1567 & $+$0.13 & ~9.8 & 8.4~~\\
UGC 2855 & SB(s)cd II-III     & 034822.6 & $+$700757 &18.7 & 1202 & $+$0.31 & 10.4 & 8.4~~\\
NGC 1569 & IBm; Sbrst	      & 043049.0 & $+$645053 &~2.5 &$-$104& $-$0.66 & ~8.7 & 8.19\\
NGC 2388 & SA(s)b: pec        & 072853.5 & $+$334905 &54.8 & 4134 & $+$1.11 & 10.9 &\nodata\\
NGC 4418 & (R$^\prime$)SAB(s)a& 122654.6 & $-$005240 &27.3 & 2179 & $+$1.12 & 10.6 &\nodata\\
NGC 6946 & SAB(rs)cd	      & 203452.3 & $+$600914 &~4.5 & ~~48 & $-$0.34 & ~9.8 & 9.28\b \\
NGC 7218 & SB(r)c	      & 221011.7 & $-$163936 &23.8 & 1662 & $-$0.17 & ~9.7 & 8.6~~\\
NGC 7771 & SB(s)a	      & 235124.8 & $+$200642 &60.0 & 4277 & $+$0.41 & 11.1 & 9.1~~\\
\enddata
\tablenotetext{a}{\footnotesize Dale et al. 2000.}
\tablenotetext{b}{\footnotesize Assuming the metallicity trend with radius found by Pilyugin \& Ferrini 1998, the targeted \HII\ region has $\log O/H+12=9.08$.}
\tablenotetext{c}{\footnotesize Metallicity references: IC~10: Lequeux et al. 1979; NGC~1569: Kobulnicky \& Skillman 1997; NGC~6946: Pilyugin \& Ferrini 1998; NGC~693, UGC~2855, NGC~7218, NGC~7771: this work}
\end{deluxetable}
\normalsize

\scriptsize
\begin{deluxetable}{llcccc}
\tablenum{2}
\label{tab:obs}
\def\a{\tablenotemark{a}}
\def\b{\tablenotemark{b}}
\def\c{\tablenotemark{c}}
\def\p{$\pm$}
\tablewidth{505pt}
\tablecaption{Palomar 200 inch Telescope Observations}
\tablehead{
\colhead{Date} &\colhead{Galaxy\a} &\colhead{R.A.~~~~~Decl.}   &\colhead{Lines\b}&
\colhead{K band} &\colhead{Sky}
\\
\colhead{}     &\colhead{}       &\colhead{(J2000)} &\colhead{}     &
\colhead{Seeing} &\colhead{Conditions} 
}
\startdata
2000 Sep 16/17 &{\bf NGC~6946}   &203452.3~$+$600914& \FeII,\Pab,\Htwo,\Brg  & 0\farcs9 & Photometric \\
               &{\bf NGC~1569~NW}&043047.3~$+$645101& \Pab~~		    & 0\farcs9 & Photometric \\
2000 Sep 17/18 &{\bf NGC~6946~NE}&203506.0~$+$601057& \FeII,\Pab,\Htwo,\Brg  & 0\farcs6 & Thin cirrus \\
               &     NGC~1569~NW &043047.3~$+$645101& \FeII~~~~~~~~~~~~~~~~~ & 0\farcs6 & Photometric \\
2000 Sep 18/19 &{\bf NGC~6946~N} &203452.3~$+$600918& \FeII,\Pab,\Htwo,\Brg  & 0\farcs7 & Thin cirrus \\
2000 Oct 13/14 &{\bf NGC~6946~S} &203452.3~$+$600910& \FeII,\Pab~~~~~~~~~~   & 0\farcs8 & Thin cirrus \\
               &     NGC~1569~NW &043047.3~$+$645101& \FeII,~~~~~~\Htwo,\Brg & 0\farcs8 & Thin cirrus \\
2000 Oct 14/15 &     NGC~6946~S  &203452.3~$+$600910& \FeII,~~~~~~\Htwo,\Brg & 0\farcs7 & Photometric \\
               &     NGC~6946    &203452.3~$+$600914& \FeII~~~~~~~~~~~~~~~~~ & 0\farcs7 & Photometric \\
               &{\bf NGC~1569~SE}&043051.7~$+$645048& ~~~~~~~~~~~~~~~~~~~\Brg& 0\farcs6 & Photometric \\
2001 Jan 06/07 &     NGC~1569~SE &043051.7~$+$645048& \Pab~~		     & 0\farcs5 & Thin cirrus \\
               &{\bf NGC~2388}   &072853.5~$+$334908& \FeII,\Pab,\Htwo,\Brg  & 0\farcs5 & Thin cirrus \\
               &{\bf NGC~4418}   &122654.6~$-$005240& \FeII,\Pab,\Htwo,\Brg  & 0\farcs6 & Thin cirrus \\
2001 Oct 03/04 &{\bf NGC~7218}   &221011.7~$-$163939& \FeII,\Pab,\Htwo,\Brg  & 0\farcs8 & Photometric \\
               &{\bf UGC~2855}   &034820.4~$+$700758& \FeII,\Pab,~~~~\Brg    & 0\farcs9 & Photometric \\
2001 Oct 04/05 &{\bf NGC~7771}   &235124.8~$+$200642& \FeII,\Pab,\Htwo,\Brg  & 0\farcs9 & Photometric \\
               &{\bf IC~10}      &002027.2~$+$591743& ~~~~~~~~\Pab,\Htwo,\Brg& 1\farcs2 & Photometric \\
2001 Oct 05/06 &     IC~10       &002027.2~$+$591743& \FeII~~~~~~~~~~~~~~~~~ & 1\farcs6\c& Thin cirrus \\
               &{\bf NGC~0693}   &015030.9~$+$060842& \FeII,\Pab,\Htwo,\Brg  & 0\farcs7 & Thin cirrus \\
2001 Nov 05/06 &     UGC~2855    &034820.4~$+$700758& ~~~~~~~~~\Htwo	     & 0\farcs8 & Photometric \\
\enddata
\tablenotetext{a}{\footnotesize Bold lettering indicates the initial observation(s) for a source.  If a source was targeted more than once, additional entries are not indicated in bold.}
\tablenotetext{b}{\footnotesize The \FeII\ line observed was either \FeIIa\ or \FeIIb, or both.}
\tablenotetext{c}{\footnotesize H band measurement.}
\end{deluxetable}
\normalsize

\scriptsize
\begin{deluxetable}{lcccccccc}
\def\a{\tablenotemark{a}}
\def\b{\tablenotemark{b}}
\def\c{\tablenotemark{c}}
\def\d{\tablenotemark{d}}
\def\e{\tablenotemark{e}}
\def\f{\tablenotemark{f}}
\tablenum{3}
\label{tab:optical_spectroscopy}
\tablewidth{535pt}
\tablecaption{Parameters Derived from Long-slit Optical Spectra}
\tablehead{
\colhead{Galaxy}                    & \colhead{Date}        & \colhead{$T_{\rm exp}$(b,r)} &
\colhead{log~(H$\beta$)\a}          &
\colhead{$c$(H$\alpha$,H$\beta$)\b} &
\colhead{log~R$_{23}$\c}            & \colhead{log~R$_3$\d} & \colhead{log~O$_{32}$\e}     &
\colhead{$\log {N \over O}$\f}
\\
\colhead{}                          & \colhead{Obs.}        & \colhead{(minutes)}          & 
\colhead{}                          & 
\colhead{}                          & \colhead{}            & 
\colhead{}                          & \colhead{}            & \colhead{}
}
\startdata
NGC 0693&Sep 1996&  40~~~~90&$-$13.6&1.05$\pm$.10&0.84$\pm$.02&$+$0.25$\pm$.02&$-$0.47$\pm$.03&$-$0.84$\pm$.06\\
UGC 2855&Sep 1997&\nodata~50&$-$14.9&2.0~$\pm$.2~&\nodata     &$-$0.26$\pm$.10&~\nodata	      &~\nodata	      \\
NGC 1569&Sep 1997&\nodata~20&$-$12.9&1.1~$\pm$.1~&\nodata     &$+$0.94$\pm$.01&~\nodata	      &~\nodata	      \\
NGC 4418&Mar 1996&  40~~~~45&\nodata&\nodata	 &\nodata     &~\nodata	      &~\nodata	      &~\nodata	      \\
NGC 7218&Sep 1996&  30~~~~90&$-$14.3&0.5~$\pm$.1~&0.69$\pm$.03&$-$0.02$\pm$.01&$-$0.62$\pm$.04&$-$0.73$\pm$.07\\
NGC 7771&Sep 1996&  60~~~~60&$-$13.8&1.1~$\pm$.1~&0.28$\pm$.10&$-$0.51$\pm$.02&$-$0.71$\pm$.12&$-$0.13$\pm$.16\\
\enddata
\tablenotetext{a}{\footnotesize  Observed flux in erg~cm$^{-2}$~s$^{-1}$, for the selected aperture.} 
\tablenotetext{b}{\footnotesize $c\equiv \log_{10} \left[ f({\rm H}\beta)_{\rm true}/f({\rm H}\beta)_{\rm obs} \right]$.  For a ratio of total-to-selective extinction of
$R = A_V/E(B-V) = 3.1, A_V = 2.1 c$.}
\tablenotetext{c}{\footnotesize The commonly-used bright-line index, R$_{23}$ = ([\ion{O}{3}] 5007$+$4959 $+$ [\ion{O}{2}] 3727)/H$\beta$; no value is listed for objects without blue data.}
\tablenotetext{d}{\footnotesize R$_3$ = ([\ion{O}{3}] 5007$+$4959)/H$\beta$.}  
\tablenotetext{c}{\footnotesize O$_{32}$ = ([\ion{O}{3}] 5007$+$4959)/[\ion{O}{2}] 3727.}
\tablenotetext{f}{\footnotesize N/O = [\ion{N}{2}] 6584/[\ion{O}{2}] 3727.}
\end{deluxetable}

\scriptsize
\begin{deluxetable}{lcrrrrrr}
\tablenum{4}
\label{tab:fluxes}
\def\a{\tablenotemark{a}}
\def\b{\tablenotemark{b}}
\def\c{\tablenotemark{c}}
\def\p{$\pm$}

\tablewidth{500pt}
\tablecaption{Line Fluxes\a\ ~and Equivalent Widths\b}
\tablehead{
\colhead{Galaxy}  & \colhead{Target}&\colhead{\hal}    &\colhead{\FeII}   &
\colhead{\Pab}    &\colhead{\FeII}  &\colhead{\Htwo}   &\colhead{\Brg}
\\
\colhead{  }      &\colhead{}       &\colhead{0.656\m} &\colhead{1.257\m} &
\colhead{1.2818\m}&\colhead{1.644\m}&\colhead{2.1218\m}&\colhead{2.1655\m}
}
\startdata
IC~10      &\HII   &1400&\nodata &200      &$<$1.5   &$<$0.7   &~17      \\
           &       &    &\nodata &190\p~~7 &$<$5.0   &$<$2.5   &~59\p~~2 \\
NGC~0693   &nucleus&~~96&\nodata &3.9	   &4.5	     &3.7      &5.4      \\
           &       &    &\nodata &1.1\p0.1 &1.8\p0.1 &2.7\p0.2 &3.7\p0.2 \\
UGC~2855   &nucleus&~~34&\nodata &~90.	   &41	     &8.9      &~31      \\
           &       &    &\nodata &~32\p~~2 &19\p~~1  &3.6\p0.2 &~24\p~~2 \\
NGC~1569~NW&\HII   &1500&6.6	 &420	   &9.1	     &6.4      &100      \\
           &       &    &1.8\p0.1&100\p~~2 &3.6\p0.2 &5.2\p0.3 &~82\p~~3 \\
NGC~1569~SE&\HII   &~400&\nodata &~55	   &\nodata  &\nodata  &~14      \\
           &       &    &\nodata &30.8\p0.7&\nodata  &\nodata  &21.9\p0.4\\
NGC~2388   &nucleus&~180&18	 &~68	   &26	     &6.7      &~12      \\
           &       &    &3.5\p0.1&18.5\p0.7&13.9\p0.8&2.8\p0.6 &5.3\p0.8 \\
NGC~4418   &nucleus&~5.5&\nodata &$<$0.5   &0.7      &16       &$<$1.1   \\
           &       &    &\nodata &$<$0.2   &0.5\p0.1 &13.9\p0.5&$<$0.9   \\
NGC~6946   &nucleus&~290&88	 &180	   &75	     &61       &~64      \\
           &       &    &2.8\p0.1&9.0\p0.2 &10.5\p0.3&6.5\p0.2 &7.0\p0.2 \\
NGC~6946~NE&\HII   &~~50&$<$1.6	 &~22	   &\nodata  &$<$2.1   &8.5      \\
           &       &    &$<$13   &~41\p~~4 &\nodata  &$<$23    &~140\p40 \\
NGC~7218   &nucleus&~~66&\nodata &8.0	   &1.9	     &$<$0.5   &1.2      \\
           &       &    &\nodata &4.8\p0.4 &1.8\p0.2 &$<$0.7   &2.4\p0.7 \\
NGC~7771   &nucleus&~160&\nodata &~40	   &12	     &26       &~15      \\
           &       &    &\nodata &7.3\p0.3 &9.7\p0.8 &16\p~~2  &5.8\p0.4 \\
\hline
$A_\lambda/A_V$\c& &0.7834&0.2903&0.2815&0.1907&0.1247&0.1204 \\
\enddata
\tablenotetext{a}{\footnotesize Fluxes corrected for atmospheric extinction are given in the first row for each galaxy and are in units of 10$^{-18}$~W~m$^{-2}$.  The \hal\ fluxes ($\sim30$\% uncertainty) are extracted from image cut-outs matched to the PIFS fields-of-view.  The absolute calibration of the near-infrared fluxes is $\sim50$\% (see text).
}
\tablenotetext{b}{\footnotesize Equivalent widths are listed in the second row for each galaxy are in units of \AA.}
\tablenotetext{c}{\footnotesize Derived from Li \& Draine 2001.}
\end{deluxetable}
\normalsize

\scriptsize
\begin{deluxetable}{lccccccc}
\tablenum{5}
\label{tab:results}
\def\a{\tablenotemark{a}}
\def\b{\tablenotemark{b}}
\def\c{\tablenotemark{c}}
\def\d{\tablenotemark{d}}
\def\e{\tablenotemark{e}}
\def\f{\tablenotemark{f}}
\def\p{$\pm$}

\tablewidth{510pt}
\tablecaption{Results}
\tablehead{
\colhead{Galaxy}&\colhead{Target}&\colhead{log ${f_\nu(6.75 \mu {\rm m}) \over f_\nu(15 \mu {\rm m})}$\a} &\colhead{$A_V$(MW,int)\b}&
\colhead{$\log N(H^o)$}&\colhead{$\log N^{\rm c}_{O8.5V}$}&
\colhead{Age$_{SB}$\d}&\colhead{Fe$^+$/H\e}
\\
\colhead{}&\colhead{}&\colhead{}&\colhead{(mag)}&
\colhead{(s$^{-1}$)} &
\colhead{} &\colhead{($10^7$~yr)}&\colhead{($10^{-7}$)}
}
\startdata
IC~10      &\HII    &$-0.59$&2.39,~0.0\p2.7   &   50.5&   1.77&   0.6  &$<$1.4 \\
NGC~0693   &nucleus &$-0.07$&0.17,11.7\p4.4   &   53.1&   4.35&   0.8  &~25    \\
UGC~2855   &nucleus &$-0.19$&2.45,~3.7\p0.8   &   53.4&   4.72&   0.7  &~26    \\
NGC~1569~NW&\HII    &$-0.93$&2.32,~0.3\p0.9   &   51.9&   3.23&   0.4  &~0.82  \\
NGC~1569~SE&\HII    &$-0.69$&2.32,~0.0\p3.8   &   51.1&   2.41&   0.5  &\nodata\\
NGC~2388   &nucleus &$-0.34$&0.19,~2.3\p0.8   &   53.8&   5.05&   0.7  &~19     \\
NGC~4418   &nucleus &$-0.79$&0.08,~~~\nodata\f&$<$52.7&$<$3.98&$>$0.9&$>$23  \\
NGC~6946   &nucleus &$-0.37$&1.13,~3.1\p0.9   &   52.3&   3.62&   0.7  &~14    \\
NGC~6946~NE&\HII    &$+0.02$&1.13,~2.8\p0.9   &   51.4&   2.69&   0.5  &$<$1.1 \\
NGC~7218   &nucleus &$-0.07$&0.11,~0.6\p0.9   &   51.9&   3.19&   0.8  &~20.   \\
NGC~7771   &nucleus &$-0.17$&0.24,~3.0\p0.9   &   53.8&   5.12&   0.7  &~12    \\
\enddata
\tablenotetext{a}{\footnotesize This mid-infrared ratio stems from flux densities extracted over a 9\arcsec\ diameter circular aperture, using images from Dale et al. 2000 that are smoothed to $\sim$9\arcsec\ resolution.}
\tablenotetext{b}{\footnotesize The foreground extinction due to the Milky Way is listed first, followed by the internal extinction.  The foreground extinctions are compiled from the Schlegel, Finkbeiner \& Davis 1998 values listed in NED, except for low Galactic latitude IC~10, for which the value is uncertain.  For IC~10, we use the value derived by Massey \& Armandroff 1995 and suggested by Hunter 2001.} 
\tablenotetext{c}{\footnotesize Derived from $\log N(H^o)$ (e.g. Vacca, Garmany, \& Shull 1996).} 
\tablenotetext{d}{\footnotesize Compiled from the instantaneous starburst curves in Leitherer \& Heckman 1995 for the \Brg\ equivalent widths.}
\tablenotetext{e}{\footnotesize The Fe$^{1+}$ gas-phase abundance is computed using Equation~\ref{eq:abundance} and assuming $T_e=5000$~K for the following line emissivities in $10^{-22}$~W~m$^{-3}$: $j$(\FeIIa)$=26.8$ and $j$(\FeIIb)$=5.94$ (Hummer et al. 1993; Pradhan \& Zhang 1993); $j({\rm Br}\gamma)=7.23\cdot10^{-7}$ applicable for the range $100<n_{\rm e} ({\rm cm}^{-3})<10,000$ (Storey \& Hummer 1995).  The uncertainties are estimated to be a factor of two.  Three targets were observed at both 1.257 and 1.644\m, and the separately-derived Fe$^{1+}$ abundances (in $10^{-7}$) are consistent with the average at this level: 0.3 vs 1.3 for NGC~1569~NW; 6.5 vs 32 for NGC~2388; 8.1 vs 20 for NGC~6946.}
\tablenotetext{f}{\footnotesize The extinction for NGC~4418 is unconstrained; the extinction based on upper limits for \Pab\ and \Brg\ is $A_V$(int)=14.4~mag.  In fact, many of the results for this peculiar system should be interpreted with caution, including the starburst ``age'' (see \S~\ref{sec:fluxes}).}
\end{deluxetable}
\normalsize

\scriptsize
\begin{deluxetable}{ccc}
\tablenum{6}
\label{tab:ratios}
\def\a{\tablenotemark{a}}
\tablewidth{320pt}
\tablecaption{Simulated Fe Ionization Fractions Fe$^{1+}$/Fe$^{2+}$/Fe$^{3+}$\a}
\tablehead{
\colhead{$T_{\rm eff}$ (K)}&\colhead{$n_{\rm e}=10$ ${\rm cm}^{-3}$}&\colhead{$n_{\rm e}=100$ ${\rm cm}^{-3}$}
}
\startdata
35,000 & 0.0793/0.755/0.166 & 0.0360/0.752/0.212\\
50,000 & 0.0746/0.218/0.707 & 0.0380/0.127/0.835\\
\enddata
\tablenotetext{a}{\footnotesize These numbers correspond to an Orion-like metallicity and 10$^{50}~{\rm s}^{-1}$ ionizing photons.}
\end{deluxetable}
\normalsize

\begin{figure}[!ht]
\epsscale{0.5}
\epsscale{1.0}
\caption[] {\ 2MASS $K$ band image (5\farcm3$\times$5\farcm3; J2000 coordinates) of IC~10 with two overlays: the PIFS field of view outlined in white and the $B$ Band 25~mag~arcsec$^{-2}$ isophote in black.} 
\label{fig:2mass1}
\end{figure}
\begin{figure}[!ht]
\epsscale{0.9}
\epsscale{1.0}
\caption[] {\ Similar to Figure~\ref{fig:2mass1} for NGC~693 and UGC~2855 (1\farcm5$\times$1\farcm5 2MASS $K$ band images).} 
\label{fig:2mass2}
\end{figure}
\begin{figure}[!ht]
\epsscale{0.9}
\epsscale{1.0}
\caption[] {\ Similar to Figure~\ref{fig:2mass1} for NGC~1569 and NGC~2388 (1\farcm5$\times$1\farcm5 2MASS $K$ band images).} 
\label{fig:2mass3}
\end{figure}
\begin{figure}[!ht]
\epsscale{0.9}
\epsscale{1.0}
\caption[] {\ Similar to Figure~\ref{fig:2mass1} for NGC~4418 and NGC~6946 (1\farcm5$\times$1\farcm5 and 6\farcm2$\times$6\farcm2 2MASS $K$ band images).} 
\label{fig:2mass4}
\end{figure}
\begin{figure}[!ht]
\epsscale{0.9}
\epsscale{1.0}
\caption[] {\ Similar to Figure~\ref{fig:2mass1} for NGC~7218 and NGC~7771 (1\farcm5$\times$1\farcm5 2MASS $K$ band images).} 
\label{fig:2mass5}
\end{figure}
\begin{figure}[!ht]
\plotone{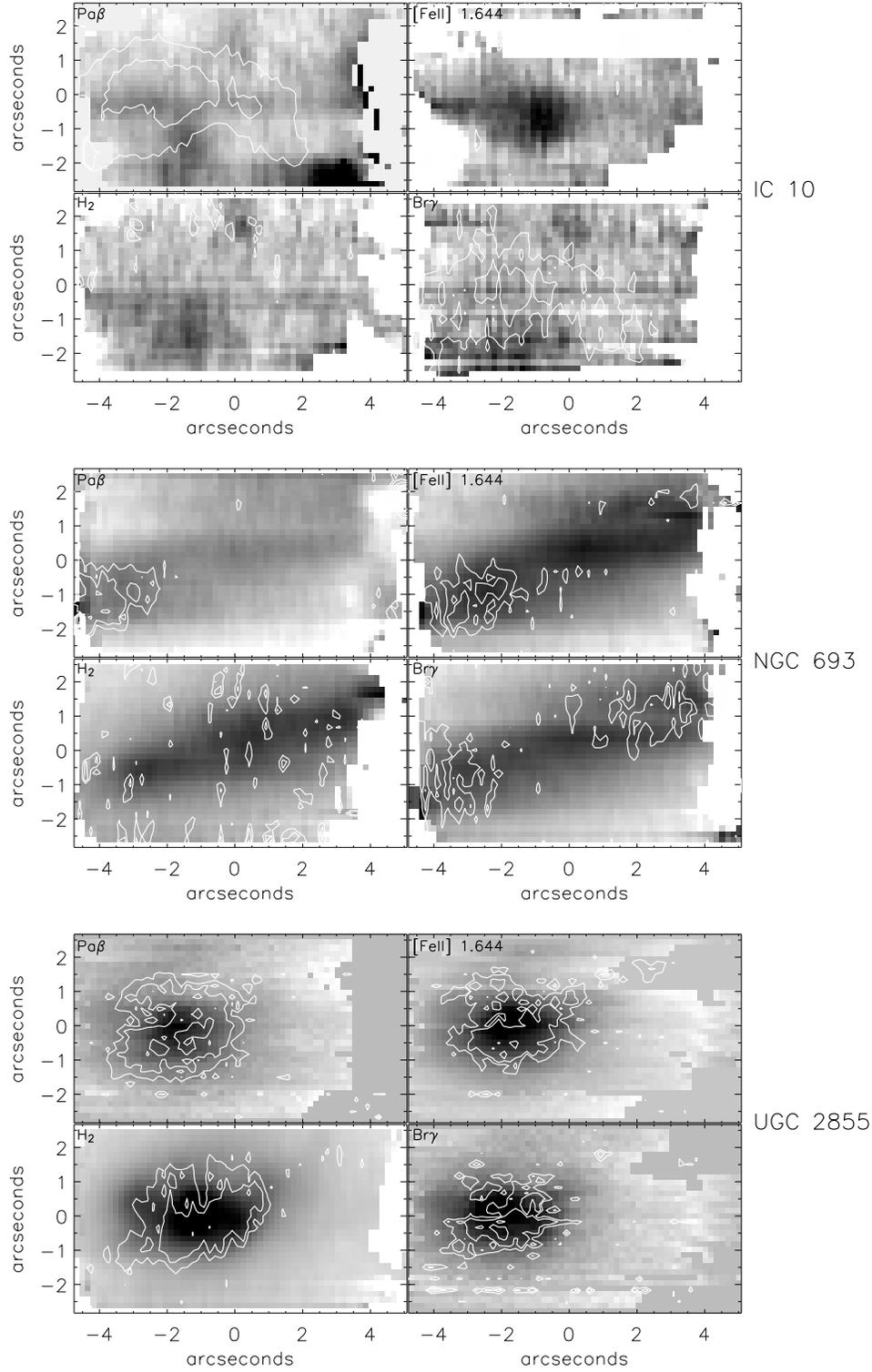}
\caption[] {\ PIFS continuum and line maps for IC~10 (top four panels), NGC~693 (middle), and UGC~2855 (bottom).  In each panel the continuum image is shown in greyscale and the line emission is portrayed as contours.  The contours are plotted at levels of 2, 3, 5, 7, 10, 14, 20, and 28$\sigma$.}
\label{fig:pifs1}
\end{figure}
\begin{figure}[!ht]
\plotone{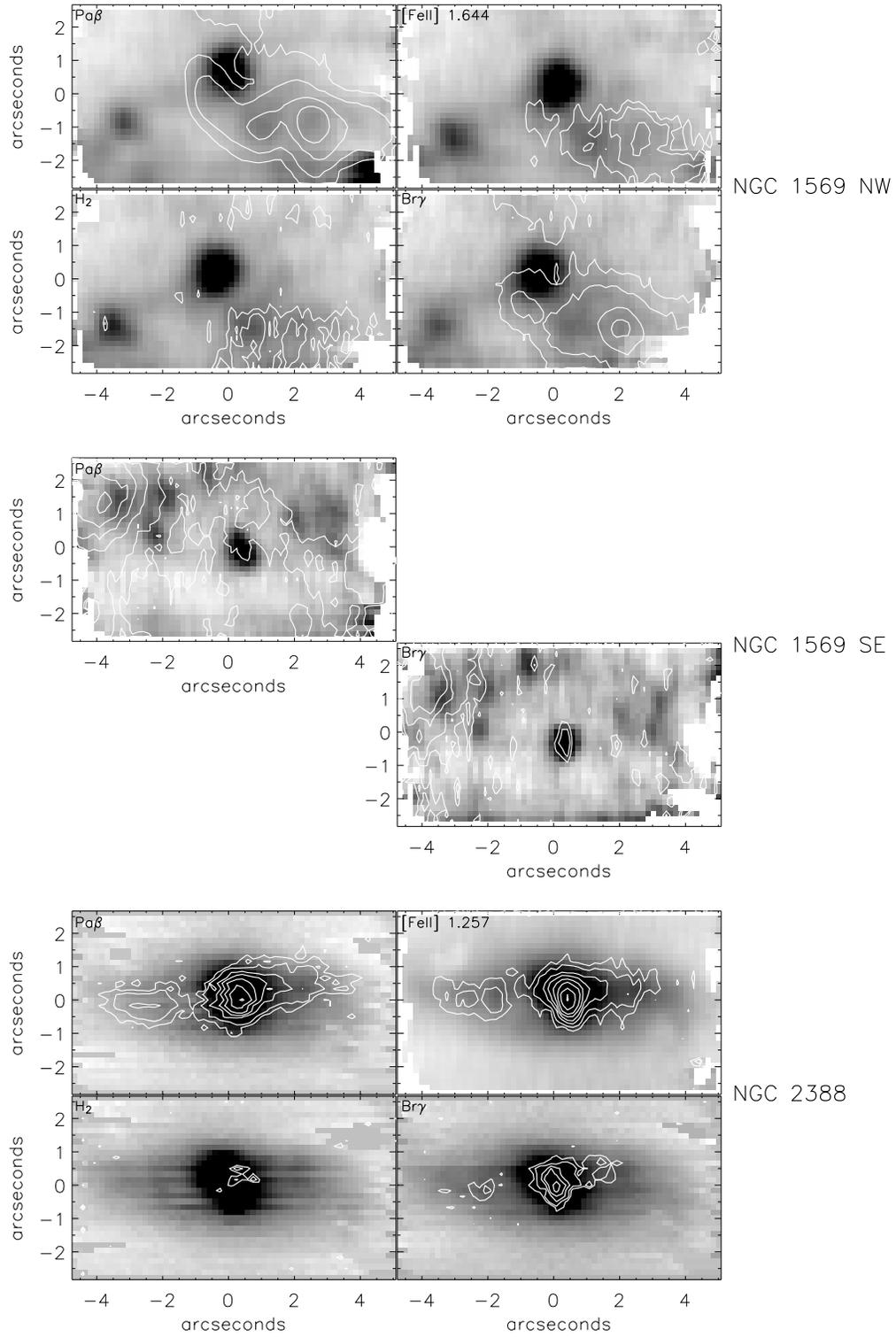}
\caption[] {\ Similar to Figure~\ref{fig:pifs1} for NGC~1569~NW (top four panels), NGC~1569~SE (middle), and NGC~2388 (bottom).}
\label{fig:pifs2}
\end{figure}
\begin{figure}[!ht]
\plotone{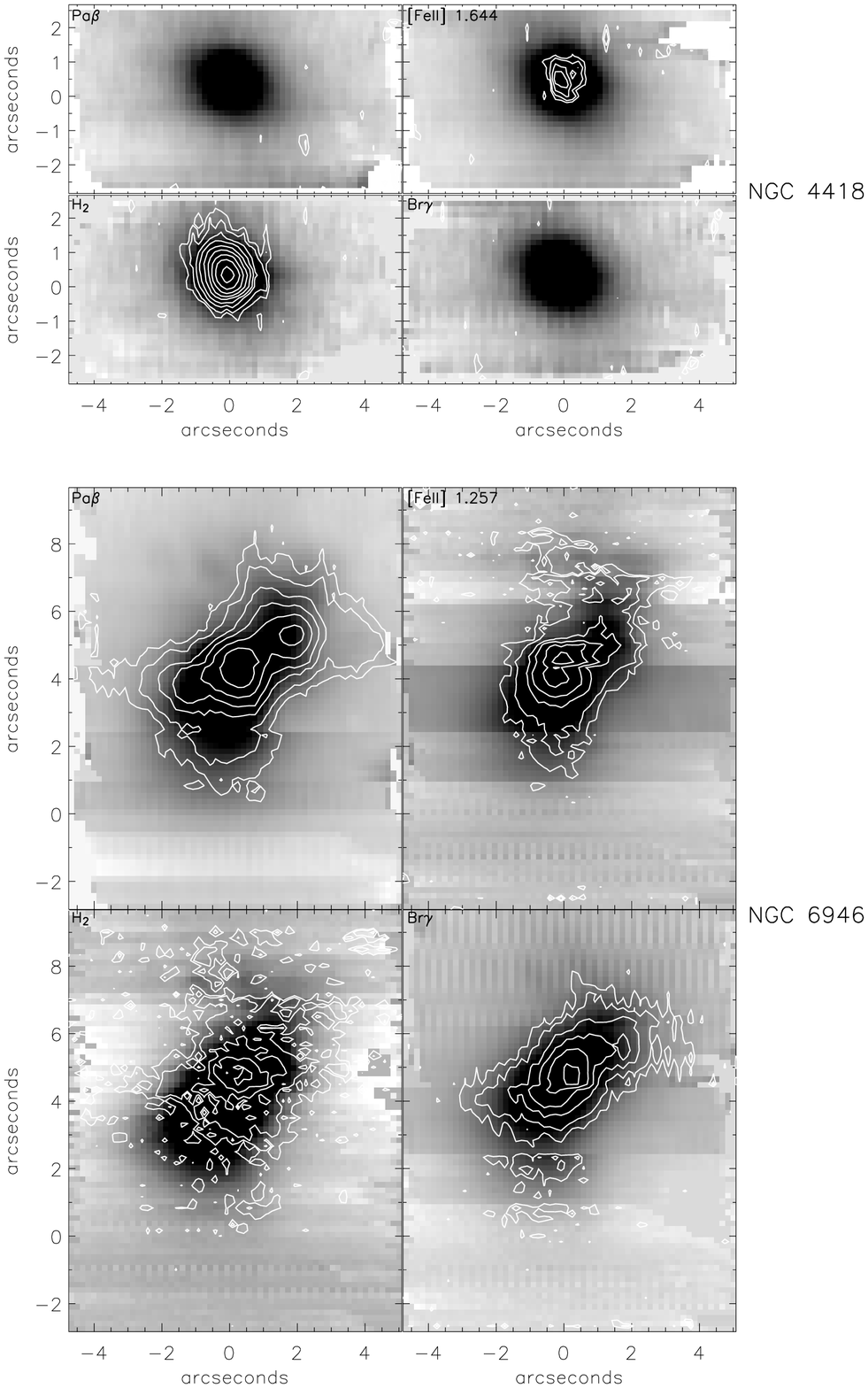}
\caption[] {\ Similar to Figure~\ref{fig:pifs1} for NGC~4418 (top four panels) and NGC~6946 (bottom).}
\label{fig:pifs3}
\end{figure}
\begin{figure}[!ht]
\plotone{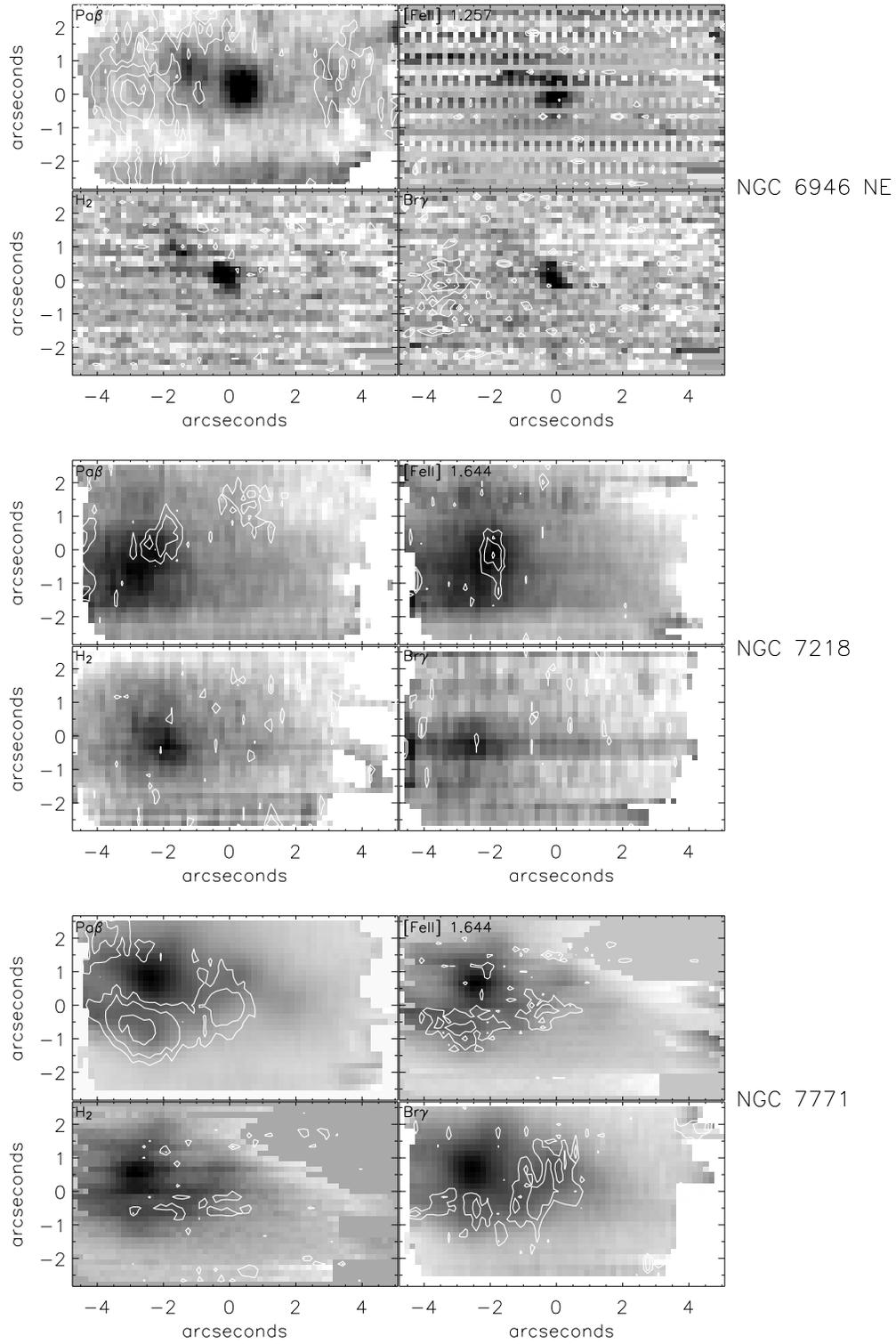}
\caption[] {\ Similar to Figure~\ref{fig:pifs1} for NGC~6946~NE (top four panels), NGC~7218 (middle), and NGC~7771 (bottom).}
\label{fig:pifs4}
\end{figure}
\clearpage
\begin{figure}[!ht]
\plotone{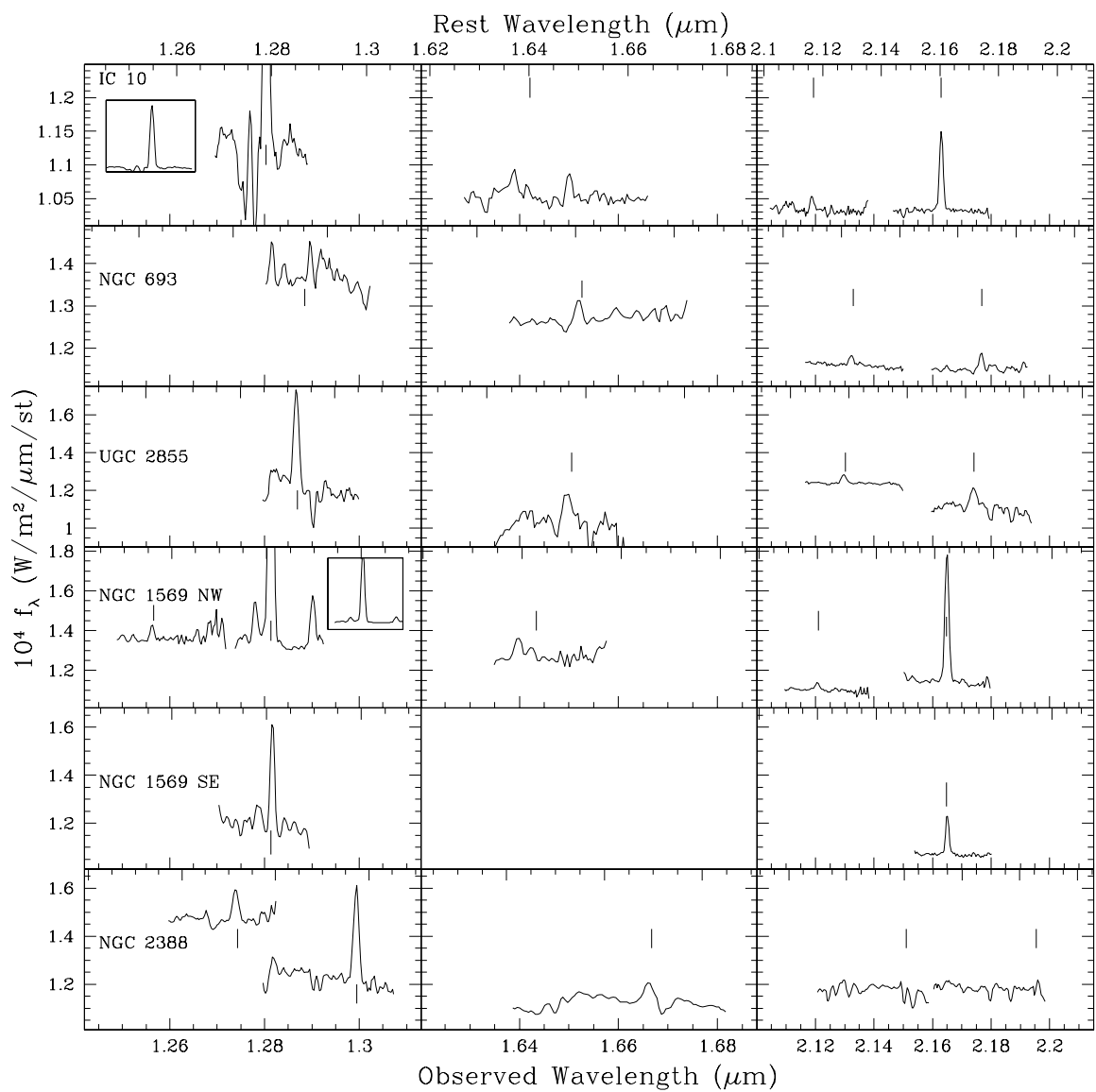}
\caption[] {\ The spectral data are plotted as the flux density surface brightness versus both rest and observed wavelength (top and bottom axes).  The lefthand panels display \FeIIa\ and \Pab\ data, the middle panels show \FeIIb\ data, and the righthand panels give the \Htwo\ and \Brg\ data.  Small vertical lines indicate the line wavelengths.  Small insets show the full profiles for IC~10 \Pab\ and NGC~1569~NW \Pab.} 
\label{fig:spectra1}
\end{figure}
\begin{figure}[!ht]
\plotone{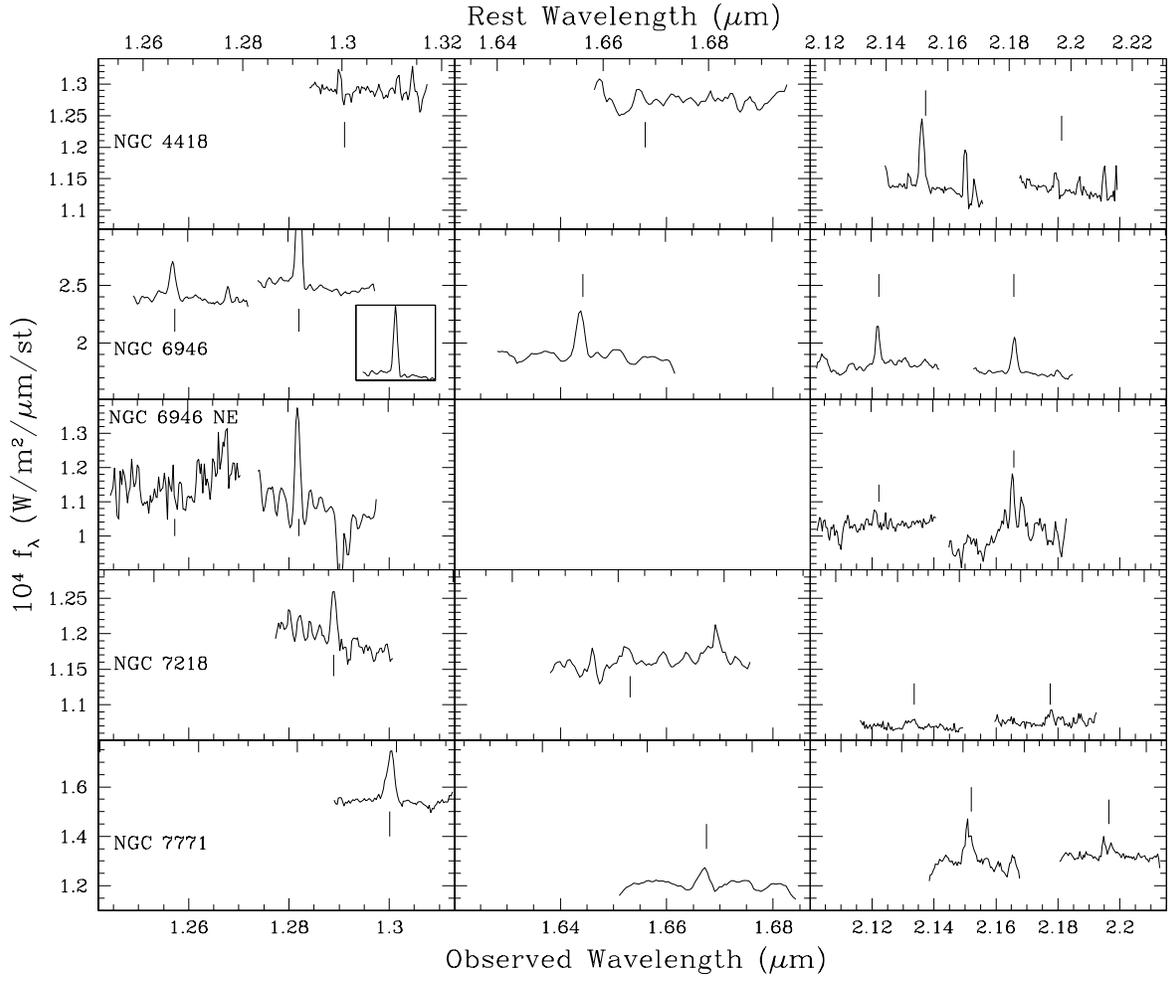}
\caption[] {\ Similar to Figure~\ref{fig:spectra1} for the remaining spectra.  A small inset shows the full profile for NGC~6946 \Pab.} 
\label{fig:spectra2}
\end{figure}
\begin{figure}[!ht]
\plotone{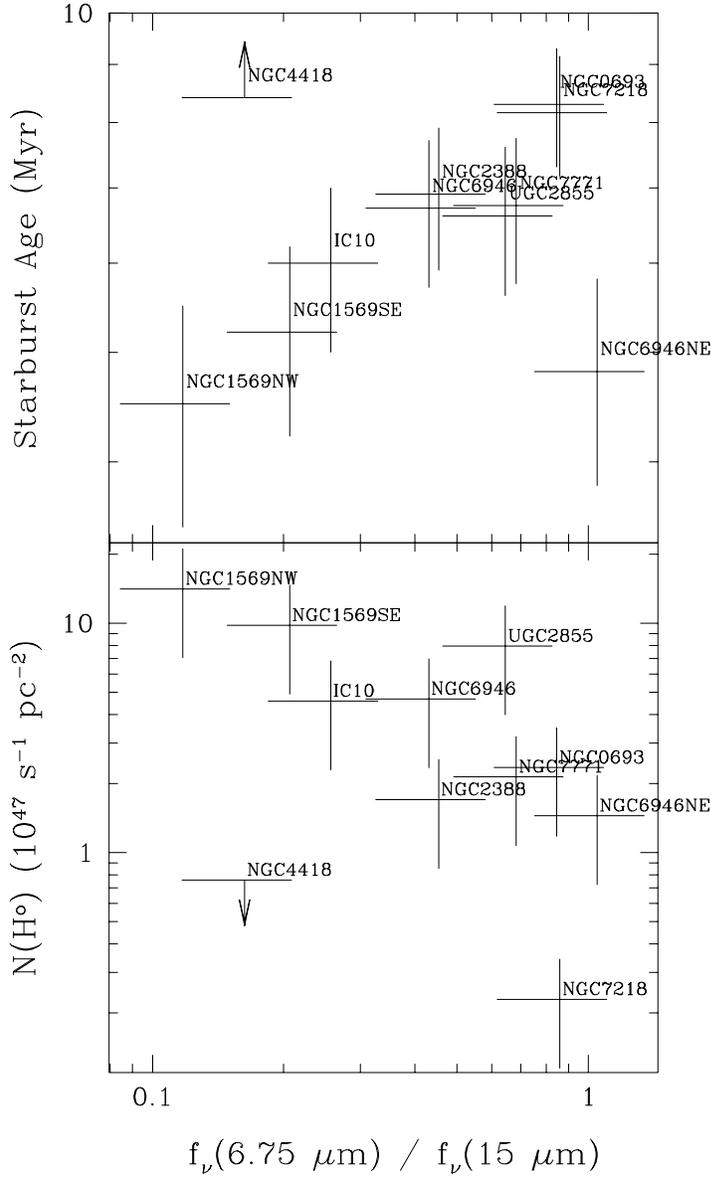}
\caption[] {\ {\bf Top}:  The starburst ages, assuming an instantaneous starburst model, as a function of mid-infrared.  {\bf Bottom}:  The density of ionizing photons strongly correlates with mid-infrared color \colore, consistent with the interpretation that lower \colore\ signifies more active star formation.} 
\label{fig:density}
\end{figure}
\begin{figure}[!ht]
\plotone{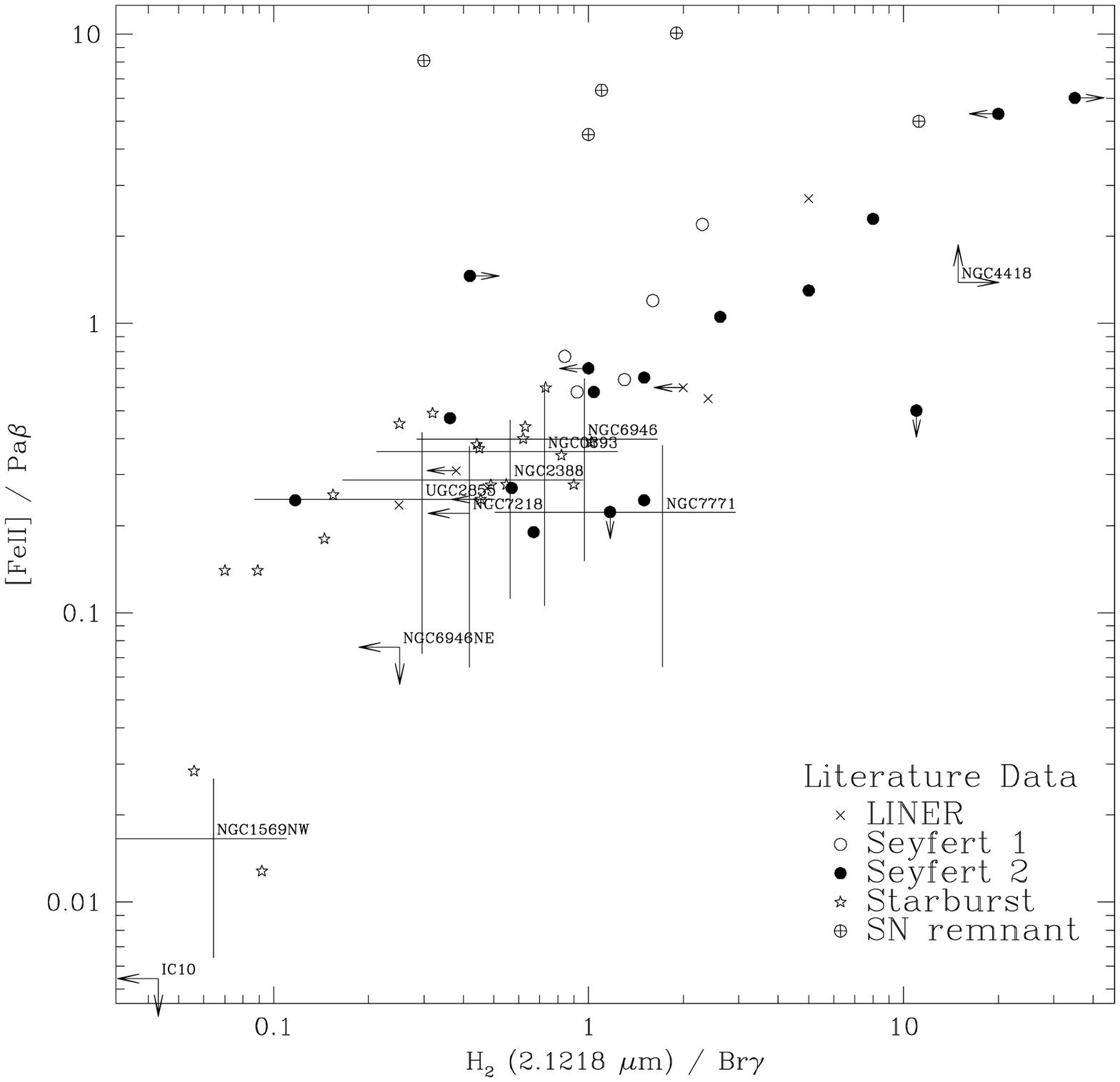}
\caption[] {\ Extinction-corrected near-infrared line ratios for this sample and for other objects drawn from the literature (Larkin et al. 1998; Calzetti 1997).  The data for NGC~4418 have not been extinction corrected due to the extinction uncertainty.  A sample of 11 Galactic compact star-forming regions spans the range $0.4\lesssim$\Htwo/\Brg$\lesssim2.2$ (Smutko \& Larkin 1999).} 
\label{fig:ratios}
\end{figure}

\newpage
\begin {thebibliography}{dum}
\bibitem[a03]{a03} Alonso-Herrero, A., Rieke, G.H., Rieke, M.J., \& Kelly, D. 2003, \aj, 125, 1210
\bibitem[b98]{b98} Baan, W.A., Salzer, J.J., \& R.D. LeWinter 1998, \apj, 509, 633 
\bibitem[b91]{b91} Baldwin, J.A., Ferland, G.J., Martin, P.G., Corbin, M.R., Cota, S.A., Peterson, B.M., \& Slettebak, A. 1991, \apj, 374, 580
\bibitem[b87]{b87} Black, J.H. \& van Dishoeck, E.F. 1987, \apj, 322, 412
\bibitem[c97]{c97} Calzetti, D. 1997, \aj, 113, 162
\bibitem[c02]{c02} Contursi, A., Kaufman, M., Helou, G., Hollenbach, D., Brahuer, J., Stacey, G., Dale, D.A., Malhotra, S., Rubio, M., Rubin, R., \& Lord, S. 2002, \aj, 124, 751
\bibitem[d99]{d99} Dale, D.A., Helou, G., Silbermann, N., Contursi, A., Malhotra, S., \& Rubin, R. 1999, \aj, 118, 2055
\bibitem[d00]{d00} Dale, D.A., Silbermann, N.A., Helou, G., Contursi, A.  et al. 2000, \aj, 120, 583
\bibitem[d01]{d01} Dale, D.A., Helou, G., Contursi, A., Silbermann, N., \& Kolhatkar, S. 2001, \apj, 549, 215
\bibitem[d02]{d02} Dale, D.A., \& Helou, G. 2002,  \apj, 159, 576
\bibitem[d97]{d97} Davies, R.I., Alonso-Herrero, A., \& Ward, M.J. 1997, \mnras, 291, 557
\bibitem[d02]{d02} Dopita, M.A., Pereira, M., Kewley, L.J., \& Capaccioli, M. 2002, \apjs, 143, 47
\bibitem[d99]{d99} Driver, S.P. 1999, \apjl, 526, L69
\bibitem[e84]{e84} Edmunds, M.G. \& Pagel, B.E.J. 1984, \mnras, 211, 507
\bibitem[e96]{e96} Engelbracht, C. W., Rieke, M.J., Rieke, G.H., \& Latter, W.B. 1996, \apj, 467, 227 
\bibitem[f01]{f01} F\"{o}rster Schreiber, N.M., Genzel, R., Lutz, D., Kunze, D., \& Sternberg, A. 2001, \apj, 552, 544
\bibitem[f98]{f98} Ferguson, A.M.N., Wyse, R.F.G., Gallagher, J.S., \& Hunter, D.A. 1998, \apjl, 506, L19  
\bibitem[g02]{g02} Garnett, D.R. 2002, \apj, 581, 1019
\bibitem[g97]{g97} Goldader, J.D., Joseph, R.D., Doyon, R., \& Sanders, D.B. 1997, \apj, 474 104
\bibitem[g95]{g95} Goldschmidt, O. \& Sternberg, A. 1995, \apj, 439, 256
\bibitem[g96]{g96} Greve, A., Becker, R., Johansson, L.E.B., \& McKeith, C.D. 1996, \aap, 312, 391
\bibitem[h99]{h99} Hamann, F. \& Ferland, G. 1999, \araa, 37, 487
\bibitem[h03]{h03} Hameed, S. \& Young, L.M. 2003, \aj, 125, 3005
\bibitem[h80]{h80} Heckman, T.M., Crane, P.C., \& Balick, B. 1980, \aap, 40, 295
\bibitem[h96]{h96} Helou, G., Malhotra, S., Beichman, C.A., Dinerstein, H., Hollenbach, D.J., Hunter, D.A., Lo, K.Y., Lord, S.D., Lu, N.Y., Rubin, R.H., Stacey, G.J., Thronson, Jr., H.A., \& Werner, M.W. 1996, \aap, 315, L157
\bibitem[h00]{h00} Helou, G., Lu, N.Y., Werner, M.W., Malhotra, S., \& Silbermann, N.A. 2000, \apjl, 532, L21
\bibitem[h01]{h01} Helou, G., Malhotra, S., Hollenbach, D., Dale, D.A., \& Contursi, A. 2001, \apjl, 548, L73
\bibitem[h97]{h97} Ho, L., Filippenko, A.V., \& Sargent, W.L. 1997, \apj, 487, 568
\bibitem[h95]{h95} Hollenbach, D. \& Natta, A. 1995, \apj, 455, 133
\bibitem[hu01]{hu01} Hunter, D., Kaufman, M., Hollenbach, D. et al. 2001, \apj, 553, 121
\bibitem[hu01]{hu01} Hunter, D. 2001, \apj, 559, 225
\bibitem[h99]{h99} H\"{u}ttemeister, S., Aalto, S., \& Wall, W.F. 1999, \aap, 346, 45
\bibitem[j00]{j00} Jansen, R.A., Fabricant, D., Franx, M., \& Caldwell, N. 2000, \apjs, 126, 331
\bibitem[k83]{k83} Kennicutt, R.C. \& Kent, S.M. 1983, \aj, 88, 1094
\bibitem[k89]{k89} Kennicutt, R.C. 1989, \apj, 344, 685 
\bibitem[k89b]{k89b} Kennicutt, R.C., Keel, W.C., \& Blaha, C.A. 1989, \aj, 97, 1022
\bibitem[k92]{k92} Kennicutt, R.C. 1992, \apj, 388, 310
\bibitem[k98]{k98} Kennicutt, R.C. 1998, \araa, 36, 189 
\bibitem[k98]{k98} Kim, D.-C. \& Sanders, D.B. 1998, \apjs, 119, 41
\bibitem[k97]{k97} Kobulnicky, H.A. \& Skillman, E.D. 1997, \apj, 489, 636
\bibitem[k99]{k99} Kobulnicky, H.A., Kennicutt, R.C., \& Pizagno, J.L. 1999, \apj, 514, 544
\bibitem[k95]{k95} Kr\"{u}ger, H., Fritze-v. Alvensleben, U., \& Loose, H.-H. 1995, \aap, 303, 41 
\bibitem[l98]{l98} Larkin, J.E., Armus, L., Knop, R.A., Soifer, B.T., \& Matthews, K. 1998, \apjs, 114, 59
\bibitem[l95]{l95} Leitherer, C. \& Heckman, T.M. 1995, \apjs, 96, 9
\bibitem[l99]{l99} Leitherer, C., Schaerer, D., Goldader, J.D., Delgado, R.M.G., Robert, C., Kune, D.F., de Mello, D.F., Devost, D., \& Heckman, T.M. 1999, \apjs, 123, 3
\bibitem[l79]{l79} Lequeux, J., Peimbert, M., Rayo, J.F., Serrano, A., \& Torres-Peimbert, S. 1979, \aap, 80, 155 
\bibitem[l94]{l94} Lequeux, J., Le Bourlot, J., Des Forets, G.P., Roueff, E., Boulanger, F., \& Rubio, M. 1994, \aap, 292, 371
\bibitem[l01]{l01} Li, A. \& Draine, B.T. 2001, \apj, 554, 778
\bibitem[l03]{l03} Lu, N., Helou, G., Werner, M.W., Dinerstein, H.L., Dale, D.A., Silbermann, N.A., Malhotra, S., Beichman, C.A., \& Jarrett, T. 2003, \apj, 588, 199
\bibitem[m01]{m01} Malhotra, S., Kaufman, M., Hollenbach, D. et al. 2001, \apj, 561, 766
\bibitem[m96]{m96} Maloney, P.R., Hollenbach, D.J., \& Tielens, A.G.G.M. 1996, \apj, 466, 561
\bibitem[m02]{m02} Martin, C.L., Kobulnicky, H.A., \& Heckman, T.M. 2002, \apj, 574, 663
\bibitem[m95]{m95} Massey, P. \& Armandroff, T.E. 1995, \aj, 109, 2470
\bibitem[m91]{m91} McGaugh, S.S. 1991, \apj, 380, 140
\bibitem[m94]{m94} McGaugh, S.S. 1994, \apj, 426, 135
\bibitem[m88]{m88} Moorwood, A.F.M. \& Oliva, E. 1988, \aap, 203, 278
\bibitem[m99]{m99} Murphy, T.W., Matthews, K., \& Soifer, B.T. 1999, \pasp, 763, 1176
\bibitem[o83]{o83} Oke, J.B. \& Gunn, J.E. 1983, \apj, 266, 713
\bibitem[p98]{p98} Pilyugin, L.S. \& Ferrini, F. 1998, \aap, 336, 103
\bibitem[p73]{p73} Pradhan, A.K. \& Zhang, H.L. 1993, \apjl, 409, L77 
\bibitem[q96]{q96} Quinet, P., Le Dourneuf, M., \& Zeippen, C.J. 1996, \aap, 120, 361
\bibitem[r00]{r00} Reunanen, J., Kotilainen, J.K., Laine, S., \& Ryder, S.D. 2000, \apj, 529, 853
\bibitem[r03]{r03} Rho, J., Reynolds, S.P., Reach, W.T., Jarrett, T.H., Allen, G.E., \& Wilson, J.C. 2003, \apj, 592, 299
\bibitem[r86]{r86} Roche, P.F., Aitken, D.K., Smith, C.H., \& James, S.D. 1986, \mnras, 218, 19
\bibitem[r03]{r03} Rodr\'\i guez, M. \& Rubin, R.H. 2003, Proceedings of IAU Symposium 217,
Sydney, Australia, July 2003, in preparation
\bibitem[ro01]{ro01} Roussel, H., Sauvage, M., Vigroux, L., Bosma, A., Bonoli, C., Gallais, P., Hawarden, T., Madden, S., \& Mazzei, P. 2001, \aap, 372, 406
\bibitem[r03]{r03} Roussel, H., Helou, G., Beck, R., Condon, J.J., Matthews, K.Y., \& Jarrett, T.H. 2003, \apj, in press
\bibitem[r91]{r91} Rubin, R.H., Simpson, J.P., Haas, M.R., \& Erickson, E.F. 1991, \pasp, 103, 834
\bibitem[r97]{r97} Rubin, R.H. et al. 1997, \apjl, 474, L131
\bibitem[s98]{s98} Sankrit, R. et al. 1998, \apj, 504, 344
\bibitem[s98]{s98} Schlegel, D.J., Finkbeiner, P.F., \& Davis, M. 1998, \apj, 500, 525
\bibitem[s99]{s99} Schlegel, E.M., Blair, W.P., \& Fesen, R.A. 2000, \aj, 120, 791 
\bibitem[sm99]{sm99} Smith, D.A., Herter, T., Haynes, M.P., \& Neff, S.G. 1999, \apj, 510, 669
\bibitem[smu99]{smu99} Smutko, M.F. \& Larkin, J.E. 1999, \apj, 117, 2448
\bibitem[s01]{s01} Spoon, H.W.W., Keane, J.V., Tielens, A.G.G.M., Lutz, D., \& Moorwood, A.F.M. 2001, \aap, 365, 353
\bibitem[s77]{s77} Stone, R.P.S. 1977, \apj, 218, 767
\bibitem[s95]{s95} Storey, P.J. \& Hummer, D.G.  1995, \mnras, 272, 41
\bibitem[t99]{t99} Taylor, C.L., H\"{u}ttemeister, S., Klein, U., \& Greve, A. 1999, \aap, 349, 424
\bibitem[t92]{t92} Terlevich, R., Tenorio-Tagle, G., France, J., \& Melnick, J. 1992, \mnras, 255, 713
\bibitem[t90]{t90} Thronson, H.A., Hunter, D.A., Casey, S., \& Harper, D.A. 1990, \apj, 355, 94
\bibitem[v96]{v96} Vacca, W.D., Garmany, C.D., \& Shull, J.M. 1996,
\apj, 460, 914
\bibitem[v00]{v00} Vanzi, L., Hunt, L.K., Thuan, T.X., \& Izotov, Y.I. 2000, \aap, 363, 493
\bibitem[v00]{v00} Volker, S. 2000, \mnras, 312, 859
\bibitem[w91]{w91} Waller, W.H. 1991, \apj, 370, 144
\bibitem[w98]{w98} Wilcots, E.M. \& Miller, B.W. 1998, \aj, 116, 2363
\end {thebibliography}
\end{document}